\documentclass[twocolumn,showpacs,preprintnumbers,superscriptaddress,amsmath,amssymb,nofootinbib]{revtex4}

\usepackage{graphicx}
\usepackage{dcolumn}
\usepackage{bm}
\usepackage{amsmath, amssymb}

\usepackage[normalem]{ulem}  
\usepackage[dvipdfmx]{color} 

\renewcommand\sout{\bgroup \color{red} \ULdepth=-.5ex \ULset}

\newcommand{\Psfig}[2]{\includegraphics[width=#1]{#2}}
\newcommand{\PsfigII}[2]{\includegraphics[scale=#1]{#2}}

\newcommand{\one}{\mbox{1}\hspace{-0.25em}\mbox{l}}

\allowdisplaybreaks

\def\naive{na\"{i}ve }
\def\naively{na\"{i}vely }
\def\Kaellen{K\"{a}llen }
\def\Schr{Schr\"{o}dinger }

\def\mev{\text{ MeV}}
\def\gev{\text{ GeV}}
\def\fm{\text{ fm}}

\def\prt{\partial}

\def\Rho{\text{P}}

\begin{document}

\preprint{}

\title{Two-body wave functions and compositeness from scattering
  amplitudes.\\ I.~General properties with schematic models}


\author{Takayasu Sekihara} 
\email{sekihara@post.j-parc.jp}
\affiliation{Advanced Science Research Center,
  Japan Atomic Energy Agency, 
  Shirakata, Tokai, Ibaraki, 319-1195, Japan}

\date{\today}

\begin{abstract}

  For a general two-body bound state in quantum mechanics, both in the
  stable and decaying cases, we establish a way to extract its
  two-body wave function in momentum space from the scattering
  amplitude of the constituent two particles.  For this purpose, we
  first show that the two-body wave function of the bound state
  corresponds to the residue of the off-shell scattering amplitude at
  the bound state pole.  Then, we examine our scheme to extract the
  two-body wave function from the scattering amplitude in several
  schematic models.  As a result, the two-body wave functions from the
  Lippmann--Schwinger equation coincides with that from the \Schr
  equation for an energy-independent interaction.  Of special interest
  is that the two-body wave function from the scattering amplitude is
  automatically scaled; the norm of the two-body wave function, to
  which we refer as the compositeness, is unity for an
  energy-independent interaction, while the compositeness deviates
  from unity for an energy-dependent interaction, which can be
  interpreted to implement missing channel contributions.  We also
  discuss general properties of the two-body wave function and
  compositeness for bound states with the schematic models.

\end{abstract}

\maketitle

\section{Introduction}

The wave function is one of the most fundamental quantities in quantum
mechanics, and to determine the wave function is the most important
subject in understanding the character of a quantum system.  This fact
can be seen especially in a bound state of two particles in a quantum
system.  In the nonrelativistic condition, such a system is governed
by the \Schr equation, and the wave function of the bound state is
evaluated as an eigenfunction of the Hamiltonian in the \Schr
equation, bringing a discrete eigenvalue, which is nothing but the
eigenenergy of the bound state.  The wave function of the bound state
represents the behavior of the two constituents inside the bound
system in coordinate or momentum space.  Namely, the squared value of
the wave function corresponds to the ``probability'' of the amplitude
of the quantum fluctuation by the constituent two particles in
coordinate or momentum space.  Moreover, the wave function can be
utilized for, {\it e.g.}, the calculation of the transition amplitude
from the bound state to other states and vice versa.

Besides, the properties of the quantum system are reflected also in
the scattering amplitude of the two particles, which is the solution
of the Lippmann--Schwinger equation.  Interestingly, if there exists a
bound state in the quantum system, the bound state must be accompanied
by a pole in the complex energy plane of the scattering amplitude for
the constituent two particles.  The pole position coincides with the
eigenvalue of the Hamiltonian in the \Schr equation associated with
the bound state wave function, and hence determining the pole position
of the bound state is equivalent to evaluating the discrete eigenvalue
of the Hamiltonian.

Then, we can \naively expect that we may extract properties of the
bound state from the scattering amplitude of two constituents,
especially from the residue of the scattering amplitude at the pole.
In this line, a famous study was done by
Weinberg~\cite{Weinberg:1965zz}.  In his study, by using the \Schr
equation and Lippmann--Schwinger equation and taking the weak binding
limit of a bound state, he expressed the scattering length and
effective range model independently in terms of the so-called field
renormalization constant, which equals unity minus norm of the
two-body wave function for the bound state.  Then, from the
experimental values of the scattering length and effective range for
the proton--neutron scattering, he concluded that the deuteron is
indeed a proton--neutron bound state.  An essential point in this
discussion is that, since the deuteron pole position exists very close
to the lower limit of the physically accessible energy, i.e., the
proton--neutron threshold, we can relate the residue of the scattering
amplitude at the deuteron pole with the observable threshold
parameters in a model independent manner.  After several decades from
the work by Weinberg, studies on the structure of near threshold bound
and resonance states have been recently done in, {\it e.g.},
Refs.~\cite{Baru:2003qq, Hyodo:2013iga, Hyodo:2014bda,
  Hanhart:2014ssa, Kamiya:2015aea, Kamiya:2016oao}.

In general cases, however, the pole position is not located near the
two-body threshold but largely below the threshold or has negatively
large imaginary part.  In such a case, we have to employ a model to
investigate the structure of the bound/resonance state.  In this line,
a well-known result from decades ago is that, for a given interaction
which generates a stable bound state, one can relate the residue of
the scattering amplitude at the pole with the wave function of the
bound state~\cite{Gottfried:2003}.  The discussion was extended in
Ref.~\cite{Hernandez:1984zzb} especially to resonance states, where
the authors proved that, with a general energy-independent
interaction, a resonance wave function in momentum space can be
obtained from the residue of the scattering amplitude of the
constituent two particles at the pole position and is correctly
normalized as unity.  Then, the structure of the bound state from the
coupled-channels scattering amplitude is investigated with a separable
interaction in Refs.~\cite{Gamermann:2009uq, YamagataSekihara:2010pj},
where the two-body wave function was found to be proportional to the
residue of the scattering amplitude at the resonance pole.  Recently,
the norm of the two-body wave function, which is called
compositeness~\cite{Hyodo:2011qc, Hyodo:2013nka}, has been extracted
from the hadron--hadron scattering amplitude with the separable
interaction for candidates of hadronic molecules in, {\it e.g.},
Refs.~\cite{Hyodo:2013nka, Aceti:2012dd, Sekihara:2012xp, Xiao:2012vv,
  Aceti:2014ala, Aceti:2014wka, Nagahiro:2014mba, Sekihara:2014kya,
  Garcia-Recio:2015jsa, Guo:2016wpy, Sekihara:2015gvw, Guo:2015daa}.

In the present study, from a more general point of view, we
investigate the relation between the wave function of the two-body
bound state in momentum space, including the case of an unstable
state, and the scattering amplitude of the two particles.  Compared to
the works in the literature, we further introduce the energy
dependence for a general two-body interaction, consider the
semirelativistic formulation and the coupled-channels problems, and
take into account self-energy effect for an unstable constituent.  We
show that solving the Lippmann--Schwinger equation at the pole
position of the bound state is equivalent to evaluating the two-body
wave function of the bound state in momentum space which is
automatically scaled.  An interesting finding is that the
compositeness from the scattering amplitude deviates from unity for
energy dependent interactions, which can be interpreted as missing
channel contributions.  A basic idea of our approach was partly given
in Ref.~\cite{Sekihara:2015gvw}, and we extend this to resonances in
practical problems, for which we employ the complex scaling
method~\cite{Aoyama:2006}.

This paper is organized as follows.  In Sec.~\ref{sec:2}, we formulate
the two-body wave functions and compositeness for bound states in
general quantum systems in the nonrelativistic and semirelativistic
conditions.  In the formulation, we show that the two-body wave
function of the bound state, both in the stable and decaying cases,
appears in the residue of the off-shell scattering amplitude at its
pole.  Next, in Sec.~\ref{sec:3}, we give numerical calculations of
the two-body wave functions and compositeness in several schematic
models for bound states.  Section~\ref{sec:4} is devoted to the
summary of this study.  This is the first paper of a series for the
two-body wave function and compositeness in general quantum systems;
our scheme constructed here will be applied to the physical $N^{\ast}$
and $\Delta ^{\ast}$ resonances in a precise scattering
amplitude~\cite{Sekihara:2016}.

\section{Formulation}
\label{sec:2}

In this section, we formulate the two-body wave function for a bound
state, regardless of whether the bound state is stable or not.  For
this purpose, we first give a setup of the quantum system in
Sec.~\ref{sec:2A}.  Next, we formulate the \Schr equation for the
bound state in Sec.~\ref{sec:2B}.  In this section we also define the
so-called compositeness as the norm of the two-body wave function for
the bound state.  Then, in Sec.~\ref{sec:2C}, we clarify how the
two-body wave function appears in the scattering amplitude, which is a
solution of the Lippmann--Schwinger equation, and propose a way to
extract the two-body wave function for the bound state from the
scattering amplitude.  The model dependence of the compositeness is
discussed in Sec.~\ref{sec:2D}.  Finally, in order to investigate
numerically the structure of a resonance state, we show our formulae
of the wave function, compositeness, and so on, in the complex scaling
method in Sec.~\ref{sec:2E}.

\subsection{Setup of the system}
\label{sec:2A}

In this paper we consider a two-body to two-body coupled-channels
scattering system.  The system is governed by the full Hamiltonian
$\hat{H}$, which can be decomposed into the free Hamiltonian
$\hat{H}_{0}$ and the interaction part $\hat{V}$:
\begin{equation}
\hat{H} = \hat{H}_{0} + \hat{V} ( E ) . 
\end{equation}
Here, we assume that we have only the two-body states in the practical
model space, i.e., we do not treat one-body bare states nor states
composed of more than two particles acted by $\hat{H}$, $\hat{H}_{0}$,
and $\hat{V}$.  In addition, we assume that the scattering process
with the interaction $\hat{V}$ is time-reversal invariant and, for the
later applications, we allow the interaction to depend intrinsically
on the energy of the system $E$, which corresponds to the eigenenergy
of the full Hamiltonian.  We neglect the spin of the scattering
particles for simplicity.

As an eigenstate of the free Hamiltonian $\hat{H}_{0}$, we introduce
the $j$th channel two-body scattering state with relative
three-momentum $\bm{q}$ as $| \bm{q}_{j} \rangle$:
\begin{equation}
\hat{H}_{0} | \bm{q}_{j} \rangle = 
\mathcal{E}_{j} ( q ) | \bm{q}_{j} \rangle , 
\quad 
\langle \bm{q}_{j} | \hat{H}_{0} = 
\mathcal{E}_{j} ( q ) \langle \bm{q}_{j} | ,
\end{equation}
where $q \equiv | \bm{q} |$ is the magnitude of the momentum $\bm{q}$.
For the eigenenergy $\mathcal{E}_{j} ( q )$, we employ two options.
One is the nonrelativistic form containing the threshold energy
\begin{equation}
\mathcal{E}_{j} ( q ) 
\equiv m_{j} + M_{j} + \frac{q^{2}}{2 \mu _{j}} ,
\quad 
\mu _{j} \equiv \frac{m_{j} M_{j}}{m_{j} + M_{j}} ,
\label{eq:Ej_NR}
\end{equation}
and the other is the semirelativistic form
\begin{equation}
\mathcal{E}_{j} ( q ) 
\equiv \sqrt{q^{2} + m_{j}^{2}} + \sqrt{q^{2} + M_{j}^{2}} .
\label{eq:Ej_SR}
\end{equation}
Here, $m_{j}$ and $M_{j}$ are masses of scattering particles in the
channel $j$ and $\mu _{j}$ is the reduced mass for them.  In the
present study, the scattering state is normalized as
\begin{equation}
\langle \bm{q}_{j}^{\prime} | \bm{q}_{k} \rangle 
= ( 2 \pi )^{3} \delta _{j k} \delta ( \bm{q}^{\prime} - \bm{q} ) .
\end{equation}
In terms of the eigenstates $| \bm{q}_{j} \rangle$, the free Hamiltonian
can be expressed as
\begin{equation}
  \hat{H}_{0} = \sum _{j} \int \frac{d^{3} q}{( 2 \pi )^{3}}
  \mathcal{E}_{j} ( q ) | \bm{q}_{j} \rangle \langle \bm{q}_{j} | .
  \label{eq:H0_dec}
\end{equation}

For the interaction $\hat{V}$, on the other hand, we employ a general
form in the following coupled-channels expression with the two-body
scattering states in coordinate space $| \bm{r} _{j} \rangle$:
\begin{equation}
  \langle \bm{r}_{j}^{\prime} | \hat{V} ( E ) | \bm{r} _{k} \rangle
  = V_{j k} ( E ; \, \bm{r}^{\prime} , \, \bm{r} ) .
  \label{eq:Vjk_coordinate}
\end{equation}
Here, $\bm{r}$ is the relative distance between two particles in the
considering channel, $r$ is its magnitude, $r \equiv | \bm{r} |$, and
the $j$th channel scattering state $| \bm{r}_{j} \rangle$ is
normalized as
\begin{equation}
  \langle \bm{r}_{j}^{\prime} | \bm{r} _{k} \rangle
  = \delta _{j k} \delta ( \bm{r}^{\prime} - \bm{r} ) .
\end{equation}
The interaction term in coordinate space $V_{j k} ( E ; \,
\bm{r}^{\prime} , \, \bm{r} )$ may, as we have mentioned, depend
intrinsically on the energy of the system $E$.  Since we assume the
time-reversal invariance of the scattering process, the interaction
term satisfies a relation
\begin{equation}
  V_{j k} ( E ; \, \bm{r}^{\prime}, \, \bm{r} )
  = V_{k j} ( E ; \, \bm{r}, \, \bm{r}^{\prime} ) ,
\end{equation}
with an appropriate choice of phases of the states.

We also consider the matrix element of the interaction with the
scattering states in momentum space:
\begin{equation}
  \langle \bm{q}_{j}^{\prime} | \hat{V} ( E ) | \bm{q}_{k} \rangle
  = \tilde{V}_{j k} ( E ; \, \bm{q}^{\prime} , \, \bm{q} ) .
  \label{eq:V_mom}
\end{equation}
This is related to the matrix element in coordinate
space~\eqref{eq:Vjk_coordinate} via the Fourier transformation as
\begin{align}
  \tilde{V}_{j k} ( E ; \, \bm{q}^{\prime} , \, \bm{q} )
  & = \int d^{3} r^{\prime} \int d^{3} r \,
  \langle \bm{q}_{j}^{\prime} | \bm{r}_{j}^{\prime} \rangle
  \langle \bm{r}_{j}^{\prime} | \hat{V} ( E ) | \bm{r}_{k} \rangle
  \langle \bm{r}_{k} | \bm{q}_{k} \rangle
  \notag \\
  & = \int d^{3} r^{\prime} \int d^{3} r \,
  V_{j k} ( E ; \, \bm{r}^{\prime} , \, \bm{r} ) e^{- i \bm{q}^{\prime} \cdot
    \bm{r}^{\prime} + i \bm{q} \cdot \bm{r}} ,
\end{align}
where we have used $\langle \bm{q}_{j} | \bm{r}_{k} \rangle = \delta
_{j k} e^{- i \bm{q} \cdot \bm{r}}$.  We note that the time-reversal
invariance of the scattering process brings a relation
\begin{equation}
  \hat{V}_{j k} ( E ; \, \bm{q}^{\prime} , \, \bm{q} )
  = \hat{V}_{k j} ( E ; \, \bm{q} , \, \bm{q}^{\prime} ) ,
\end{equation}
with an appropriate choice of phases of the states.

\subsection{\Schr equation}
\label{sec:2B}

Let us now suppose that the full Hamiltonian has a discrete eigenstate
$| \psi _{L M} \rangle$ in the partial wave $L$ with its azimuthal
component $M$, in addition to the ordinary scattering states.  We
formulate the \Schr equation and wave function for this discrete
eigenstate.  If its eigenvalue $E_{\rm pole}$ is a real number, we can
treat the $| \psi _{L M} \rangle$ state as a usual stable bound state.
On the other hand, if $E_{\rm pole}$ has an imaginary part, the
eigenstate $| \psi _{L M} \rangle$ is a resonance state; $\text{Re} \,
E_{\rm pole}$ and $- 2 \, \text{Im} \, E_{\rm pole}$ are the mass and
width of the resonance state, respectively.  Since we treat both bound
and resonance states on the same footing in the following discussions,
we formulate the \Schr equation and wave function in a manner
applicable to both cases.  In any case, the state $| \psi _{L M}
\rangle$ is a solution of the \Schr equation
\begin{equation}
\hat{H} | \psi _{L M} \rangle 
= \left [ \hat{H}_{0} + \hat{V} ( E_{\rm pole} ) \right ] | \psi _{L M} \rangle
= E_{\rm pole} | \psi _{L M} \rangle .
\label{eq:Schr_op}
\end{equation}
Then, to establish the normalization of the resonance state, we should
employ the Gamow vector~\cite{Hernandez:1984zzb, Gamow:1928zz,
  Hokkyo:1965zz, Berggren:1968zz, Romo:1968zz}, where we take $\langle
\tilde{\psi} _{L M} | \equiv \langle \psi _{L M}^{\ast} |$ instead of
$\langle \psi _{L M} |$ for the bra vector of the resonance.  In this
notation, we can normalize the resonance wave function in the
following manner:
\begin{equation}
\langle \tilde{\psi} _{L M^{\prime}} | \psi _{L M} \rangle 
= \delta _{M^{\prime} M} .
\label{eq:norm_unity}
\end{equation}
The \Schr equation for the resonance bra state is expressed with the
same eigenenergy as
\begin{equation}
\langle \tilde{\psi} _{L M} | \hat{H} 
= \langle \tilde{\psi} _{L M} 
| \left [ \hat{H}_{0} + \hat{V} ( E_{\rm pole} ) \right ]
= \langle \tilde{\psi} _{L M} | E_{\rm pole} .
\end{equation}

From the \Schr equation in the operator form~\eqref{eq:Schr_op}, we
can formulate the usual \Schr equation with the c-number wave
function.  Since in this study we formulate the \Schr equation in
momentum space and solve it, we employ the $j$th channel wave function
in momentum space $\tilde{\psi} _{j} ( \bm{q} ) \equiv \langle
\bm{q}_{j} | \psi _{L M} \rangle$.  By using the expressions of
$\hat{H}_{0}$ and $\hat{V}$ in Eqs.~\eqref{eq:H0_dec} and
\eqref{eq:V_mom}, we can express the \Schr equation~\eqref{eq:Schr_op}
as
\begin{align}
  & \mathcal{E}_{j} ( q ) \tilde{\psi} _{j} ( \bm{q} )
  + \sum _{k} \int \frac{d^{3} q^{\prime}}{( 2 \pi )^{3}}
  \tilde{V}_{j k} ( E_{\rm pole} ; \, \bm{q} , \, \bm{q}^{\prime} )
  \tilde{\psi} _{k} ( \bm{q}^{\prime} )
  \notag \\ 
  & = E_{\rm pole} \tilde{\psi} _{j} ( \bm{q} ) ,
  \label{eq:Schr_NR}
\end{align}
where we have inserted a relation
\begin{equation}
  \one _{\rm model} = 
  \sum _{k} \int \frac{d^{3} q^{\prime}}{( 2 \pi )^{3}} 
  | \bm{q}_{k}^{\prime} \rangle \langle \bm{q}_{k}^{\prime} | ,
  \label{eq:one_model}
\end{equation}
which is valid in the practical model space, between $\hat{V}$ and $|
\psi _{L M} \rangle$.  This \Schr equation can be simplified with the
partial wave decomposition.  Namely, on the one hand, the interaction
term can be decomposed as
\begin{equation}
  \tilde{V}_{j k} ( E ; \, \bm{q}^{\prime} , \, \bm{q} )
  = \sum _{L = 0}^{\infty} ( 2 L + 1 ) 
V_{L, j k} ( E ; \, q^{\prime} , \, q ) 
P_{L} ( \hat{q}^{\prime} \cdot \hat{q} ) ,
\label{eq:Vdec}
\end{equation}
with $\hat{q} \equiv \bm{q} / q$ being the direction of the vector
$\bm{q}$, and each partial wave component can be calculated as
\begin{equation}
V_{L, j k} ( E ; \, q^{\prime}, \, q )
= \frac{1}{2} \int _{-1}^{1} d ( \hat{q}^{\prime} \cdot \hat{q} )
P_{L} ( \hat{q}^{\prime} \cdot \hat{q} )
\tilde{V}_{j k} ( E ; \, \bm{q}^{\prime} , \, \bm{q} ) ,
\end{equation}
thanks to the relation for the Legendre polynomials $P_{L} ( x )$
\begin{equation}
\int _{-1}^{1} d x P_{L} ( x ) P_{L^{\prime}} ( x ) 
= \frac{2}{2 L + 1} \delta _{L L^{\prime}} .
\label{eq:PL_norm}
\end{equation}
On the other hand, the wave function $\tilde{\psi} _{j} ( \bm{q} )$
consists of the radial part $R_{j} ( q )$ and the spherical harmonics
$Y_{L M} ( \hat{q} )$ as
\begin{equation}
  \tilde{\psi} _{j} ( \bm{q} ) = R_{j} ( q ) Y_{L M} ( \hat{q} ) .
  \label{eq:psi_RY}
\end{equation}
We fix the normalization of the spherical harmonics $Y_{L M} ( \hat{q}
)$ as
\begin{equation}
\int d \Omega _{\bm{q}} Y_{L M} ( \hat{q} )
Y_{L^{\prime} M^{\prime}}^{\ast} ( \hat{q} ) 
= 4 \pi \delta _{L L^{\prime}} \delta _{M M^{\prime}} ,
\label{eq:Y_norm}
\end{equation}
with the solid angle $\Omega _{\bm{q}}$ for $\bm{q}$.  By using the
above expressions in the partial wave decomposition, we can rewrite the
\Schr equation~\eqref{eq:Schr_NR} as
\begin{align}
  & \mathcal{E}_{j} ( q ) R_{j} ( q )
  + \sum _{k} \int _{0}^{\infty} \frac{d q^{\prime}}{2 \pi ^{2}}
  q^{\prime 2} V_{L, j k} ( E_{\rm pole} ; \, q, \, q^{\prime} ) R_{k} ( q^{\prime} )
  \notag \\
  & = E_{\rm pole} R_{j} ( q ) ,
  \label{eq:Schr_final}
\end{align}
where we have used the relation in Eq.~\eqref{eq:Y_norm} and
\begin{equation}
  P_{L} ( \hat{q}^{\prime} \cdot \hat{q} )
  = \frac{1}{2 L + 1} \sum _{M = - L}^{L} 
  Y_{L M} ( \hat{q}^{\prime} ) Y_{L M}^{\ast} ( \hat{q} ) .
  \label{eq:PYY}
\end{equation}
The \Schr equation~\eqref{eq:Schr_final} is the final form to evaluate
the radial part of the two-body wave function $R_{j} ( q )$ for stable
bound states.  For resonance states, we employ the complex scaling
method, which is explained in Sec.~\ref{sec:2E}.  We note that we
solve the integral equation~\eqref{eq:Schr_final} numerically by
discretizing the momentum and replacing the integral with respect to
the momentum with a summation.

Before going to the formulation of the Lippmann--Schwinger equation,
we here comment on the norm of the two-body wave function for the
resonance state.  Namely, while the wave function in momentum space
can be written as in Eq.~\eqref{eq:psi_RY} from the ket state $| \psi
_{L M} \rangle$, the two-body wave function from the bra state
$\langle \tilde{\psi} _{L M} |$, $\langle \tilde{\psi} _{L M} |
\bm{q}_{j} \rangle$, can be evaluated as
\begin{equation}
\langle \tilde{\psi} _{L M} | \bm{q}_{j} \rangle 
= \langle \psi _{L M}^{\ast} | \bm{q}_{j} \rangle 
= R_{j} ( q ) Y_{L M}^{\ast} ( \hat{q} ) .
\label{eq:psiq-ast}
\end{equation}
Here we emphasize that, while we take the complex conjugate for the
spherical harmonics, we do not take for the radial part.  This is
because, while the spherical part can be calculated and normalized in
a usual sense, the radial part should be treated so as to remove the
divergence of the wave function at $r \to \infty$ when we calculate
the norm (see discussions in Refs.~\cite{Hernandez:1984zzb,
  Hyodo:2013nka, Hokkyo:1965zz, Berggren:1968zz, Romo:1968zz}).  From
the above wave function, we can calculate the norm with respect to the
$j$th channel two-body wave function, $X_{j}$, in the following
manner:
\begin{equation}
X_{j} \equiv \int \frac{d^{3} q}{( 2 \pi )^{3}} 
\langle \tilde{\psi} _{L M} | \bm{q}_{j} \rangle 
\langle \bm{q}_{j} | \psi _{L M} \rangle 
= \int _{0}^{\infty} d q \, \Rho _{j} ( q ) , 
\end{equation}
where we have introduced a density distribution $\Rho _{j} ( q )$:
\begin{equation}
\Rho _{j} ( q ) \equiv \frac{q^{2}}{2 \pi ^{2}} 
\left [ R_{j} ( q ) \right ] ^{2} .
\label{eq:norm}
\end{equation}
The quantity $X_{j}$ is referred to as the compositeness.  We note
that the compositeness $X_{j}$ as well as the wave function
$\tilde{\psi} _{j} ( \bm{q} )$ is not observable and hence in general
a model dependent quantity (see the discussion in Sec.~\ref{sec:2D}).
As we can see from the construction, for the resonance state, the
compositeness $X_{j}$ is given by the complex number squared of the
radial part $R_{j} ( q )$ rather than by the absolute value squared,
which is essential to normalize the resonance wave function.
Therefore, in general the compositeness becomes complex for resonance
states.

We also note that the sum of the norm $X_{j}$ should be unity if there
is no missing (or implicit) channels, which would be an eigenstate of
the free Hamiltonian, to describe the bound state.  However, in actual
calculations we may have contributions from missing channels, which
can be implemented into the interaction and be origin of the intrinsic
energy dependence of the interaction.  In such a case, denoting the
missing channels representatively as $| \psi _{0} \rangle$, we can
decompose unity in terms of the eigenstates of the free Hamiltonian:
\begin{equation}
\one = | \psi _{0} \rangle \langle \psi _{0} | 
+ \sum _{j} \int \frac{d^{3} q}{( 2 \pi )^{3}} 
| \bm{q}_{j} \rangle \langle \bm{q}_{j} | ,
\end{equation}
instead of Eq.~\eqref{eq:one_model}.  Therefore, introducing the
missing channel contribution $Z$ defined as
\begin{equation}
Z \equiv  \langle \tilde{\psi} _{L M} 
| \psi _{0} \rangle \langle \psi _{0} | \psi _{L M} \rangle ,
\label{eq:missing-I}
\end{equation}
which is a model dependent quantity and becomes complex for resonance
states as well, we can express the normalization of the wave function
$| \psi _{L M} \rangle$ as
\begin{equation}
\langle \tilde{\psi} _{L M} | \psi _{L M} \rangle 
= Z + \sum _{j} X_{j} = 1 .
\label{eq:sum_rule}
\end{equation}
Thus, in contrast to the \naive manner in quantum mechanics, in the
present study we do not make the sum of the compositeness $X_{j}$
coincide with unity by hand.  Instead, as we will discuss in
Sec.~\ref{sec:2C}, the value of the norm, $X_{j}$, is automatically
fixed without any artificial factor when we calculate the residue of
the scattering amplitude.

\subsection{Lippmann--Schwinger equation}
\label{sec:2C}

In this subsection we consider the same problem as in the previous
subsection with the scattering amplitude.  In particular, we show how
we can extract the two-body wave function of the bound state from the
scattering amplitude.  Although the relation between the bound-state
wave function and the residue of the scattering amplitude at the pole
position is already discussed in the literature, in this study we
treat a more general case with an energy dependent interaction in a
coupled-channels problem.  We note that some of the formulation
presented here is already given in Ref.~\cite{Sekihara:2015gvw}, but
for completeness we give it in detail.

In general, the scattering amplitude can be formally obtained with the
Lippmann--Schwinger equation in an operator form:
\begin{align}
\hat{T} ( E ) = & \hat{V} ( E ) 
+ \hat{V} ( E ) \frac{1}{E - \hat{H}_{0}} \hat{T} ( E ) 
\notag \\
= & \hat{V} ( E ) + \hat{V} ( E ) \frac{1}{E - \hat{H}} \hat{V} ( E ) ,
\label{eq:LS}
\end{align}
with the $T$-matrix operator $\hat{T}$, the free Hamiltonian
$\hat{H}_{0}$, and the full Hamiltonian $\hat{H} \equiv \hat{H}_{0} +
\hat{V} ( E )$.  We use the same $\hat{H}$, $\hat{H}_{0}$ and
$\hat{V}$ as in the previous subsection.  From the $T$-matrix operator
$\hat{T}$, we can evaluate scattering amplitude of the $k ( \bm{q} )
\to j ( \bm{q}^{\prime} )$ scattering, where $\bm{q}^{( \prime )}$ is
the relative three-momentum in the initial (final) state, as
\begin{equation}
  T_{j k} ( E ; \, \bm{q}^{\prime} , \, \bm{q} )
  \equiv \langle \bm{q}_{j}^{\prime} | \hat{T} ( E ) | \bm{q}_{k} \rangle .
\end{equation}
The scattering state $| \bm{q}_{j} \rangle$ is again the same as in
the previous subsection.  Due to the time-reversal invariance, the
scattering amplitude satisfies
\begin{equation}
T_{j k} ( E; \, \bm{q}^{\prime} , \, \bm{q} ) 
= T_{k j} ( E ; \, \bm{q} , \, \bm{q}^{\prime} ) , 
\end{equation}
with an appropriate choice of phases of the states.  The scattering
amplitude $T_{j k} ( E ; \, \bm{q}^{\prime} , \, \bm{q})$ is a
solution of the Lippmann--Schwinger equation in the following form:
\begin{align}
& T_{j k} ( E ; \, \bm{q}^{\prime} , \, \bm{q} ) 
= \tilde{V}_{j k} ( E ; \, \bm{q}^{\prime} , \, \bm{q} ) 
\notag \\
& + \sum _{l} \int \frac{d^{3} k}{( 2 \pi )^{3}} 
\frac{\tilde{V}_{j l} ( E ; \, \bm{q}^{\prime} , \, \bm{k} ) 
T_{l k} ( E ; \, \bm{k} , \, \bm{q} )}
{E - \mathcal{E}_{l} ( k )} ,
\label{eq:LS-II}
\end{align}
where the interaction term $\tilde{V}_{j k}$ has been introduced in
Eq.~\eqref{eq:V_mom}.

Next, let us decompose the scattering amplitude into partial wave
amplitudes, as in Eq.~\eqref{eq:Vdec}:
\begin{equation}
T_{j k} ( E ; \, \bm{q}^{\prime} , \, \bm{q} ) 
= \sum _{L = 0}^{\infty} ( 2 L + 1 ) 
T_{L, j k} ( E ; \, q^{\prime} , \, q ) 
P_{L} ( \hat{q}^{\prime} \cdot \hat{q} ) ,
\end{equation}
\begin{equation}
T_{L, j k} ( E ; \, q^{\prime} , \, q ) 
= \frac{1}{2} \int _{-1}^{1} d ( \hat{q}^{\prime} \cdot \hat{q} )
P_{L} ( \hat{q}^{\prime} \cdot \hat{q} ) 
T_{j k} ( E ; \, \bm{q}^{\prime} , \, \bm{q} ) .
\label{eq:PWA_dec}
\end{equation}
Since the Legendre polynomials satisfy the following relation
\begin{equation}
\int d \Omega _{\bm{k}} P_{L} ( \hat{q}^{\prime} \cdot \hat{k} )
P_{L^{\prime}} ( \hat{k} \cdot \hat{q} )
= \frac{4 \pi}{2 L + 1} \delta _{L L^{\prime}} 
P_{L} ( \hat{q}^{\prime} \cdot \hat{q} ) ,
\end{equation}
we can rewrite the Lippmann--Schwinger equation~\eqref{eq:LS-II} as
\begin{align}
& T_{L, j k} ( E ; \, q^{\prime} , \, q ) 
= V_{L, j k} ( E ; \, q^{\prime} , \, q ) 
\notag \\
& + \sum _{l} \int _{0}^{\infty} \frac{d k}{2 \pi ^{2}} k^{2}
\frac{V_{L, j l} ( E ; \, q^{\prime} , \, k ) 
T_{L, l k} ( E ; \, k , \, q )}
{E - \mathcal{E}_{l} ( k )} .
\label{eq:LS_final}
\end{align}
This is the final form of the scattering amplitude to be solved for
stable bound states.  For resonance states we employ the complex
scaling method explained in Sec.~\ref{sec:2E} to calculate the
scattering amplitude.  In the numerical calculations of the integral
equation~\eqref{eq:LS_final}, we discretize the momentum so as to
replace the integral with a summation.

Here we comment on the relation between the energy $E$ and momenta $q$
and $q^{\prime}$.  In the physical scattering, the initial and final
states should be on their mass shell and their energy should be fixed
as $E = \mathcal{E}_{j} ( q^{\prime} ) = \mathcal{E}_{k} ( q )$.  We
call the scattering amplitude in this condition as the on-shell
amplitude.  An important feature is that, for open channels, the
on-shell scattering amplitude is in general observable and can be
determined model independently in a partial wave analysis.  The
on-shell scattering amplitude in each partial wave satisfies the
optical theorem from the unitarity of the $S$-matrix, and in our
formulation its expression is:
\begin{equation}
\text{Im} \, T_{L, j j} ( E )^{\text{on-shell}}
= - \sum _{k} 
\frac{\rho _{k} ( E )}{2}
\left | T_{L, j k} ( E )^{\text{on-shell}}
\right | ^{2} ,
\end{equation}
where the sum runs over the open channels and $\rho _{j} ( E )$ is the
phase space in channel $j$, whose expression is
\begin{equation}
  \rho _{j} ( E )
  \equiv \frac{\mu_{j} k_{j} ( E )}{\pi} ,
  \quad
  k_{j} ( E ) \equiv \sqrt{2 \mu _{j} (E - m_{j} - M_{j})} ,
  \label{eq:kj_NR}
\end{equation}
for the nonrelativistic case~\eqref{eq:Ej_NR}, and 
\begin{equation}
  \rho _{j} ( E )
  \equiv \sqrt{k_{j} ( E )^{2} + m_{j}^{2}}
  \sqrt{k_{j} ( E )^{2} + M_{j}^{2}}\frac{k_{j} ( E )}{\pi E},
\end{equation}
\begin{equation}
  k_{j} ( E )
  \equiv \frac{\lambda ^{1/2} ( E^{2}, \, m_{j}^{2}, \, M_{j}^{2} )}
         {2 E},
  \label{eq:kj_SR}
\end{equation}
for the semirelativistic case~\eqref{eq:Ej_SR} with the \Kaellen
function $\lambda (x, \, y, \, z) = x^{2} + y^{2} + z^{2} - 2 x y - 2
y z - 2 z x$.

Furthermore, one can perform the analytic continuation of the on-shell
amplitude to the complex energy and to the pole position for the bound
state, keeping the relation $E = \mathcal{E}_{j} ( q^{\prime} ) =
\mathcal{E}_{k} ( q )$ with complex momenta $q$ and $q^{\prime}$.
Therefore, the pole position $E_{\rm pole}$ and residue of the
on-shell amplitude at this pole are in principle determined only with
the experimental quantities in a model independent manner.

However, in contrast to the on-shell amplitude, we can mathematically
treat the scattering amplitude $T_{L, j k} ( E ; \, q^{\prime} , \,
q)$ as a function of three independent variables $E$, $q^{\prime}$,
and $q$ as an off-shell amplitude.  In particular, we can perform the
analytic continuation of the scattering amplitude by taking complex
value of the energy $E$ but keeping the momenta $q$ and $q^{\prime}$
real and positive, with which the relation $E = \mathcal{E}_{j} (
q^{\prime} ) = \mathcal{E}_{k} ( q )$ is no longer satisfied.  This
condition of the complex energy $E$ but real and positive momenta $q$
and $q^{\prime}$ will be essential to extract the wave function from
the scattering amplitude at the pole position of the bound state in
the complex energy plane.

We now explain the way to extract the two-body wave function and
compositeness from the off-shell scattering amplitude obtained by the
analytic continuation for the energy.  The key is the factor $1 / (E -
\hat{H})$ in the Lippmann--Schwinger equation~\eqref{eq:LS}.  We start
with the fact that the off-shell scattering amplitude as well as the
on-shell one has the bound state pole at $E = E_{\rm pole}$.
Actually, near the pole, the off-shell scattering amplitude is
dominated by the pole term in the expansion by the eigenstates of the
full Hamiltonian, and hence we have
\begin{equation}
\hat{T} ( E ) \approx \sum _{M = - L}^{L} 
\hat{V} ( E_{\rm pole} ) | \psi _{L M} \rangle \frac{1}{E - E_{\rm pole}}
\langle \tilde{\psi} _{L M} | \hat{V} ( E_{\rm pole} ) ,
\label{eq:Tapprox}
\end{equation}
where we have summed up the possible azimuthal component $M$.  In this
expression, the operator $| \psi _{L M} \rangle \langle \tilde{\psi}
_{L M} |$ coincides with a projector of rank $1$ attached to the
resonance pole in Ref.~\cite{Guo:2015daa}.  This fact indicates the
proper normalization of the bound-state wave function $\langle
\tilde{\psi} _{L M} | \psi _{L M} \rangle = 1$.\footnote{In terms of
  the propagator for the bound state, the normalization of the bound
  state vector, $\langle \tilde{\psi} _{L M} | \psi _{L M} \rangle =
  1$, is guaranteed by the relation
  \begin{equation}
    \frac{1}{E - \hat{H} ( E )} \approx
    | \psi _{L M} \rangle \frac{1}{E - E_{\rm pole}}
    \langle \tilde{\psi}_{L M} | ,
  \end{equation}
  around the pole position $E = E_{\rm pole}$, which is the basis of
  Eq.~\eqref{eq:Tapprox}.  Actually, in the right-hand side of the
  above equation, the field-renormalization constant for the bound
  state, which coincides with the residue of the bound-state
  propagator, is chosen to be exactly unity.}  Then, it is important
that the wave function $| \psi _{L M} \rangle$ appears in the residue
of the scattering amplitude at the pole.  Evaluating the matrix
element of this $T$-matrix operator, we obtain
\begin{align}
& T_{j k} ( E ; \, \bm{q}^{\prime} , \, \bm{q} ) 
\notag \\
& \approx \sum _{M = - L}^{L} 
\frac{\langle \bm{q}_{j}^{\prime} | \hat{V} ( E_{\rm pole} )
  | \psi _{L M} \rangle 
  \langle \tilde{\psi} _{L M} | \hat{V} ( E_{\rm pole} ) | \bm{q}_{k} \rangle}
     {E - E_{\rm pole}} .
\end{align}
The matrix elements in the numerator of the above expression, $\langle
\bm{q}_{j}^{\prime} | \hat{V} ( E_{\rm pole} ) | \psi _{L M} \rangle$
and $\langle \tilde{\psi} _{L M} | \hat{V} ( E_{\rm pole} ) |
\bm{q}_{k} \rangle$, can be calculated as follows.  By using the \Schr
equation~\eqref{eq:Schr_op}, the former matrix element is calculated
as
\begin{align}
\langle \bm{q}_{j} | \hat{V} ( E_{\rm pole} ) | \psi _{L M} \rangle
= & \langle \bm{q}_{j} | \left ( \hat{H} - \hat{H}_{0} \right ) 
| \psi _{L M} \rangle 
\notag \\
= & \left [ E_{\rm pole} - \mathcal{E}_{j} ( q ) \right ] 
\langle \bm{q}_{j} | \psi _{L M} \rangle , 
\end{align}
and from Eq.~\eqref{eq:psi_RY} we obtain
\begin{equation}
\langle \bm{q}_{j} | \hat{V} ( E_{\rm pole} ) | \psi _{L M} \rangle
= \gamma _{j} ( q ) Y_{L M} ( \hat{q} ) , 
\end{equation}
where we have defined $\gamma _{j} ( q )$ as
\begin{equation}
\gamma _{j} ( q ) \equiv 
\left [ E_{\rm pole} - \mathcal{E}_{j} ( q ) \right ] 
R_{j} ( q ) .
\label{eq:gamma}
\end{equation}
In a similar manner we can calculate the latter matrix element as
\begin{equation}
\langle \tilde{\psi} _{L M} | \hat{V} ( E_{\rm pole} ) | \bm{q}_{j} \rangle
= \gamma _{j} ( q ) Y_{L M}^{\ast} ( \hat{q} ) .
\end{equation}
Therefore, by using the above matrix elements, we can rewrite the
scattering amplitude near the pole as
\begin{align}
T_{j k} ( E ; \, \bm{q}^{\prime} , \, \bm{q} ) 
& \approx 
\frac{\gamma _{j} ( q^{\prime} ) \gamma _{k} ( q )}{E - E_{\rm pole}} 
\sum _{M = - L}^{L} 
Y_{L M} ( \hat{q}^{\prime} ) Y_{L M}^{\ast} ( \hat{q} ) 
\notag \\
& = ( 2 L + 1 )
\frac{\gamma _{j} ( q^{\prime} ) \gamma _{k} ( q )}{E - E_{\rm pole}} 
P_{L} ( \hat{q}^{\prime} \cdot \hat{q} ) ,
\label{eq:Tapprox-II}
\end{align}
where we have used the formula in Eq.~\eqref{eq:PYY}.  Since the
amplitude in Eq.~\eqref{eq:Tapprox-II} is nothing but the $L$-wave
component, we can show that indeed the $L$-wave partial wave amplitude
contains the pole:
\begin{equation}
T_{L, j k} ( E ; \, q^{\prime} , \, q ) = 
\frac{\gamma _{j} ( q^{\prime} ) \gamma _{k} ( q )}{E - E_{\rm pole}}
+ (\text{regular at } E = E_{\rm pole}) .
\end{equation}

Then, interestingly, the residue of the partial wave amplitude $\gamma
_{j} ( q )$ contains information on the two-body wave function as in
Eq.~\eqref{eq:gamma}.  Actually, we can calculate the $j$th channel
compositeness, $X_{j}$, by using the residue $\gamma _{j} ( q )$ as
\begin{equation}
  X_{j} = \int _{0}^{\infty} d q \, \Rho _{j} ( q ) ,
  \quad
  \Rho _{j} ( q ) \equiv \frac{q^{2}}{2 \pi ^{2}} 
  \left [ \frac{\gamma _{j} ( q )}
    {E_{\rm pole} - \mathcal{E}_{j} ( q )} \right ] ^{2} .
\label{eq:X_gamma}
\end{equation}
This is the formula to evaluate the $j$th channel compositeness
$X_{j}$ from the residue of the partial wave amplitude $T_{L}$ at the
pole.  The residue $\gamma _{j} ( q )$ is extracted as a function of
the real and positive momentum $q$ from the off-shell amplitude with
complex energy $E \to E_{\rm pole}$.

We emphasize that the scattering amplitude, and hence its residue
$\gamma _{j} ( q )$, is obtained from the Lippmann--Schwinger equation
without introducing any extra factor to scale the value of the
compositeness $X_{j}$, since the Lippmann--Schwinger equation is an
inhomogeneous integral equation.  This means that the value of the
compositeness in Eq.~\eqref{eq:X_gamma} as well as that of the
two-body wave function is automatically fixed without any scaling
factor when we calculate it from the residue of the scattering
amplitude.  In other words, the compositeness from the scattering
amplitude is uniquely determined once we fix the model space, form of
the kinetic energy $\mathcal{E}_{j}(q)$, and interaction.

The fact that the value of the compositeness in Eq.~\eqref{eq:X_gamma}
is automatically fixed leads to the question on the normalization of
the compositeness.  Actually, a single-channel problem with an
energy-independent interaction in a general form was discussed in
Ref.~\cite{Hernandez:1984zzb} and the authors found that the norm of
the two-body wave function from the scattering amplitude is exactly
unity in this case.  However, in general the sum of the compositeness
from the scattering amplitude may deviate from unity.  Therefore, we
can define the rest part of the normalization of the total wave
function~\eqref{eq:norm_unity} $Z$ as
\begin{equation}
  Z \equiv 1 - \sum _{j} X_{j} .
  \label{eq:missing-II}
\end{equation}
Since the compositeness $X_{j}$ is defined as the norm of the two-body
wave function, $Z$ can be interpreted as the missing-channel
contribution.  In Sec.~\ref{sec:3} we will see that the
missing-channel contribution $Z$ is exactly zero for the
energy-independent interaction, but $Z$ becomes nonzero if we switch
on the energy dependence of the interaction.

Finally we comment on the probabilistic interpretation of the
compositeness for resonances.  As we have discussed, for resonances
the compositeness from each channel is in general complex.  This means
that, strictly speaking, we cannot interpret them as the probability
to find each channel component inside the resonance state.  In order
to cure this point, we would like to introduce quantities
$\tilde{X}_{j}$ and $\tilde{Z}$ by following the discussion in
Ref.~\cite{Sekihara:2015gvw} as
\begin{equation}
  \tilde{X}_{j} \equiv \frac{ | X_{j} | }{1 + U} ,
  \quad
  \tilde{Z} \equiv \frac{ | Z | }{1 + U} ,
  \label{eq:XtildeZtilde}
\end{equation}
with 
\begin{equation}
  U \equiv \sum _{j} | X_{j} | + | Z | - 1 .
  \label{eq:XtildeU}
\end{equation}
The quantities $\tilde{X}_{j}$ and $\tilde{Z}$ are real, bound in the
range $[0, \, 1]$, and automatically satisfy the sum rule:
\begin{equation}
  \sum _{j} \tilde{X}_{j} + \tilde{Z} = 1 .
  \label{eq:Xtilde_sum}
\end{equation}
The quantity $U$ measures how much the imaginary part of the
compositeness $X_{j}$ or missing-channel contribution $Z$ is
nonnegligible and/or their real part is largely negative.  In
particular, with $U \ll 1$ one can expect a similarity between the
resonance state considered and a wave function of a stable bound
state.  In this sense, if and only if $U \ll 1$, we can interpret
$\tilde{X}_{j}$ ($\tilde{Z}$) as the probability of finding the
composite (missing) part.

Besides, there are several ways to interpret the complex compositeness
such as taking $\text{Re} \, X$~\cite{Aceti:2014ala}, $| X
|$~\cite{Aceti:2012dd, Guo:2016wpy, Guo:2015daa}, and $( 1 + |X| - |Z|
)/ 2$~\cite{Kamiya:2015aea, Kamiya:2016oao}.  In principle these
quantities may deviate from $\tilde{X}$, but, in practice, when we
treat narrow resonances, one can expect $U \ll 1$, with which we will
obtain similar results in any approaches: $\text{Re} \, X \simeq | X |
\simeq ( 1 + |X| - |Z| )/ 2 \simeq \tilde{X}$.

\subsection{Model dependence of the compositeness}
\label{sec:2D}

Here we discuss the model dependence of the compositeness.

First, as we know, the wave function $\tilde{\psi}_{j} ( q )$ and
two-body interaction are not observable and hence model dependent
quantities.  In the present approach, the model dependence of the wave
function can be understood with the property of the residue $\gamma
_{j} ( q )$, which becomes a factor of the wave function as in
Eq.~\eqref{eq:gamma}.  Namely, while the on-shell scattering
amplitude, including its analytic property for the complex energy, is
in principle determined in a model independent manner, the off-shell
amplitude is not observable.  Therefore, in order to evaluate the
residue $\gamma _{j} ( q )$ of the off-shell amplitude with real
momentum $q$ and complex energy $E = E_{\rm pole}$, some model which
fixes $\tilde{V}_{j k} ( E ; \, \bm{q}^{\prime} , \, \bm{q} )$ as a
function of three independent variables $E$, $\bm{q}^{\prime}$, and
$\bm{q}$ is in general necessary.

One may expect that the integrated quantity of the wave function,
i.e., the compositeness, is model independent.  However, as
demonstrated in Ref.~\cite{Nagahiro:2014mba}, the compositeness $X$ is
also not observable even for stable bound states because the field
renormalization constant for the bound state $Z = 1 - X$ is not a
physical observable.  This situation is similar to that of the
deuteron $d$-wave probability $P_{D}$, which is known as not
observable~\cite{Machleidt:2000ge}.

However, when the pole exists very close to the threshold of interest,
one can express the compositeness only with the observable quantities
in a model independent manner, as done in Refs.~\cite{Weinberg:1965zz,
  Baru:2003qq, Hyodo:2013iga, Hanhart:2014ssa, Hyodo:2014bda,
  Kamiya:2015aea, Kamiya:2016oao}.  In this case, we can expand
quantities in powers of $(E - E_{\rm pole})/E_{\rm typ}$, where
$E_{\rm typ}$ is a typical energy scale of the system, around $E =
E_{\rm pole}$.  In particular, assuming that the pole position of an
$s$-wave bound state of $j$th channel is very close to its threshold,
we have~\cite{Weinberg:1965zz}
\begin{equation}
  \langle \bm{q}_{j} | \hat{V} ( E ) | \psi _{0 0} \rangle =
  \langle \tilde{\psi} _{0 0} | \hat{V} ( E ) | \bm{q}_{j} \rangle =
  g_{0} + \mathcal{O} \left ( \frac{E - E_{\rm pole}}{E_{\rm typ}} \right ) ,
\end{equation}
where $g_{0}$ is a constant.  Then, through the constant $g_{0}$, one
can directly relate the compositeness and threshold parameters in the
scattering amplitude such as the scattering length and effective range
regardless of the details of the interaction~\cite{Weinberg:1965zz}.

Besides, in a certain model the wave function and compositeness can be
determined only with quantities of the on-shell amplitude.  Actually,
we may consider a separable interaction that is a function only of the
energy $E$ in momentum space,
\begin{equation}
  \langle
  \bm{q}_{j}^{\prime} | \hat{V} ( E ) | \bm{q}_{k} \rangle =
  \mathcal{V}_{j k} ( E ) ,
  \label{eq:Vseparable}
\end{equation}
which is valid, {\it e.g.}, if the interaction range is negligible
compared to the typical length of the system.  For this interaction,
the scattering amplitude can be evaluated in an algebraic equation as
\begin{equation}
  T ( E )
  = \left [ \mathcal{V} ( E )^{-1} - \mathcal{G} ( E ) \right ] ^{-1} ,
\end{equation}
with the two-body loop function\footnote{Regularization is necessary
  to tame the ultraviolet divergence of the integral in
  $\mathcal{G}$.}
\begin{equation}
  \mathcal{G}_{j} ( E ) = \int \frac{d^{3} k}{( 2 \pi )^{3}}
  \frac{1}{E - \mathcal{E}_{j} ( k )} .
\end{equation}
An important feature in this prescription is that the off-shell
scattering amplitude coincides with the on-shell one.  Therefore, one
can evaluate the residue of the off-shell scattering amplitude at the
pole position from the analytic continuation of the on-shell
scattering amplitude.\footnote{The residue of the on-shell
  scattering amplitude, $\gamma _{j}^{\prime}$,
  often called the coupling constant, is related to the residue of the
  off-shell amplitude $\gamma _{j} ( q )$ as
  \begin{equation}
    \gamma _{j}^{\prime} = \gamma _{j} ( k_{\rm on} ) ,
  \end{equation}
  with the ``on-shell'' momentum $k_{\rm on}$ determined with
  Eq.~\eqref{eq:kj_NR} or \eqref{eq:kj_SR} with the energy $E = E_{\rm
    pole}$.}  This fact indicates that with the
interaction~\eqref{eq:Vseparable} one can describe compositeness only
with empirical quantities of the on-shell amplitude.  However, we
should emphasize that, although the compositeness is expressed only
with model-independent quantities, this does not mean that the
compositeness is a model independent quantity.  Indeed, employing the
separable interaction~\eqref{eq:Vseparable} is nothing but choosing
one model for the interaction.

\subsection{Formulae in the complex scaling method for resonances}
\label{sec:2E}

In numerical calculations of the two-body wave function for resonance
states, we have to treat the complex eigenvalue $E_{\rm pole}$ both in
the \Schr equation and Lippmann--Schwinger equation.  In particular,
the complex eigenvalue $E_{\rm pole}$ exists in the second
(unphysical) Riemann sheet of the complex energy plane.  In order to
perform the numerical calculation in this condition, we employ the
complex scaling method in this study.  The details of the complex
scaling method are given in, {\it e.g.}, Ref.~\cite{Aoyama:2006}.  In
this subsection, we briefly show the formulae of the wave function,
compositeness, and so on in the complex scaling method.

In the complex scaling method, we transform the relative coordinate
$\bm{r}$ and relative momenta $\bm{q}$ into the complex-scaled value
in the following manner:
\begin{equation}
  \bm{r} \to \bm{r} e^{i \theta} ,
  \quad 
  \bm{q} \to \bm{q} e^{- i \theta} ,
\end{equation}
with a certain positive angle $\theta$.  An important fact is that,
with the complex scaling for the momentum, we can reach the second
Riemann sheet of the complex energy plane.  In this sense, the angle
$\theta$ should be large enough to go to the resonance pole position
$E_{\rm pole}$.  We also note that the angle $\theta$ has an upper
limit in order to maintain the convergence of integrals with a complex
variable.

With this transformation, the scattering state $| \bm{q}_{j} \rangle$
becomes $| \bm{q}_{j} e^{- i \theta} \rangle$, and hence the
eigenvalue of the free Hamiltonian $\hat{H}_{0}$ and the wave function
$\langle \bm{q} _{j} | \psi _{L M} \rangle$ respectively become
\begin{equation}
  \begin{split}
    & \hat{H}_{0} | \bm{q}_{j} e^{- i \theta} \rangle = 
    \mathcal{E}_{j} ( q e^{- i \theta} ) | \bm{q}_{j} e^{- i \theta} \rangle , 
    \\ 
    & \langle \bm{q}_{j} e^{- i \theta} | \hat{H}_{0} = 
    \mathcal{E}_{j} ( q e^{- i \theta} ) \langle \bm{q}_{j} e^{- i \theta} | ,
  \end{split}
\end{equation}
and
\begin{equation}
  \begin{split}
    & \langle \bm{q}_{j} e^{- i \theta} | \psi _{L M} \rangle
    = R_{j} ( q e^{- i \theta} ) Y_{L M} ( \hat{q} ),
    \\
    & \langle \tilde{\psi} _{L M} | \bm{q}_{j} e^{- i \theta} \rangle
    = R_{j} ( q e^{- i \theta} ) Y_{L M}^{\ast} ( \hat{q} ),
  \end{split}
\end{equation}
where we note that the spherical harmonics stays unchanged, since only
the behavior of the radial part is relevant to the convergence of the
resonance wave function.

Then the complex-scaled \Schr equation~\eqref{eq:Schr_final} can be
expressed as
\begin{align}
  & \mathcal{E}_{j} ( q e^{- i \theta} ) R_{j} ( q e^{- i \theta} )
  \notag \\
  & + e^{- 3 i \theta} \sum _{k} \int _{0}^{\infty}
  \frac{d q^{\prime}}{2 \pi ^{2}} q^{\prime 2}
  V_{L, j k} ( E ; \, q e^{- i \theta} , \, q^{\prime} e^{- i \theta} )
  R_{k} ( q^{\prime} e^{- i \theta} )
  \notag \\
  & = E_{\rm pole} R_{j} ( q e^{- i \theta} ) .
  \label{eq:Schr_CSM}
\end{align}
Here we emphasize that the eigenenergy $E_{\rm pole}$ is stable with
respect to the change of the angle $\theta$, which is in contrast to
the scattering state, which scales with the $q e^{- i \theta}$
dependence for the momentum $q$.  The norm of the wave function can be
calculated by
\begin{equation}
  X_{j} = \int _{0}^{\infty} d q \, \Rho _{j} ( q ) ,
  \label{eq:Xj_CSM_Schr}
\end{equation}
with the complex-scaled density distribution
\begin{equation}
  \Rho _{j} ( q ) \equiv e^{- 3 i \theta} \frac{q^{2}}{2 \pi ^{2}} 
       \left [ R_{j} ( q e^{- i \theta} ) \right ] ^{2},
\label{eq:norm_CSM}
\end{equation}
where the factor $e^{- 3 i \theta}$ has been introduced to the density
distribution $\Rho _{j} ( q )$ so as to reproduce the original formula
in Eq.~\eqref{eq:norm} when we change the integral variable as
$q^{\prime} \equiv q e^{- i \theta}$.

In a similar manner, the Lippmann--Schwinger
equation~\eqref{eq:LS_final} becomes
\begin{align}
  & T_{L, j k} ( E ; \, q^{\prime} e^{- i \theta} , \, q e^{- i \theta} ) 
  = V_{L, j k} ( E ; \, q^{\prime} e^{- i \theta} , \, q e^{- i \theta} ) 
  \notag \\
  & + e^{- 3 i \theta}
  \sum _{l} \int _{0}^{\infty} \frac{d k}{2 \pi ^{2}} k^{2}
  \frac{V_{L, j l} ( E ; \, q^{\prime} e^{- i \theta} , \, k e^{- i \theta} )}
       {E - \mathcal{E}_{l} ( k e^{- i \theta} )}
  \notag \\
  & \quad \quad \quad \quad \quad \quad \quad 
  \times T_{L, l k} ( E ; \, k e^{- i \theta} , \, q e^{- i \theta} )
  ,
\label{eq:LS_CSM}
\end{align}
where we have taken the matrix element of the $T$-matrix operator as
$\langle \bm{q}_{j}^{\prime} e^{- i \theta} | \hat{T} ( E ) |
\bm{q}_{k} e^{- i \theta} \rangle$.  The scattering amplitude $T_{L, j
  k} ( E ; \, q^{\prime} e^{- i \theta} , \, q e^{- i \theta} )$ has a
resonance pole at $E = E_{\rm pole}$, whose position is again stable
with respect to the change of the angle $\theta$.  Around the
resonance pole, the scattering amplitude is represented as
\begin{align}
  & T_{L, j k} ( E ; \, q^{\prime} e^{- i \theta} , \, q e^{- i \theta} )
  \notag \\
  & = 
  \frac{\gamma _{j} ( q^{\prime} e^{- i \theta} ) \gamma _{k} ( q e^{- i \theta} )}
       {E - E_{\rm pole}}
       + (\text{regular at } E = E_{\rm pole}) .
\end{align}
The residue $\gamma _{j} ( q e^{- i \theta} )$ can be evaluated in the
same manner to the previous subsection as
\begin{equation}
  \gamma _{j} ( q e^{- i \theta} ) \equiv 
  \left [ E_{\rm pole} - \mathcal{E}_{j} ( q e^{- i \theta} ) \right ] 
  R_{j} ( q e^{- i \theta} ) ,
  \label{eq:gamma_CSM}
\end{equation}
where we have used the following formulae in the complex scaling
method
\begin{align}
  \langle \bm{q}_{j} e^{- i \theta} | \hat{V} ( E_{\rm pole} ) | \psi _{L M} \rangle
  = & \langle \bm{q}_{j} e^{- i \theta} | \left ( \hat{H} - \hat{H}_{0} \right ) 
  | \psi _{L M} \rangle 
  \notag \\
  = & \gamma _{j} ( q e^{- i \theta} ) Y_{L M} ( \hat{q} ) , 
\end{align}
\begin{equation}
  \langle \tilde{\psi} _{L M} | \hat{V} ( E_{\rm pole} )
  | \bm{q}_{j} e^{- i \theta} \rangle
  = \gamma _{j} ( q e^{- i \theta} ) Y_{L M}^{\ast} ( \hat{q} ) .
\end{equation}
Now we can calculate the compositeness from the residue of the
scattering amplitude at the resonance pole $E_{\rm pole}$.  Actually,
taking into account the complex-scaled quantities above, we have the
density distribution as
\begin{equation}
  \Rho _{j} ( q ) \equiv e^{- 3 i \theta} \frac{q^{2}}{2 \pi ^{2}} 
    \left [ \frac{\gamma _{j} ( q e^{- i \theta} )}
      {E_{\rm pole} - \mathcal{E}_{j} ( q e^{- i \theta} )} \right ] ^{2} ,
  \label{eq:X_CSM}
\end{equation}
where the factor $e^{- 3 i \theta}$ has been introduced to the density
distribution $\Rho _{j} ( q )$ again as in Eq.~\eqref{eq:norm_CSM}.

\section{Two-body wave functions and compositeness in schematic models}
\label{sec:3}

Let us now give the numerical results on the two-body wave functions
and compositeness, i.e., the norm of the two-body wave function,
extracted from the scattering amplitude.  For this purpose, we employ
four schematic models.  The first one is a single-channel problem to
generate stable bound states in Sec.~\ref{sec:3A}.  With this model we
examine our scheme and check the normalization of the wave functions
of the bound states from the scattering amplitude.  We also discuss
how the energy dependence of the interaction affects the wave
functions and compositeness.  Then, the second model is a
single-channel problem to generate a resonance state in
Sec.~\ref{sec:3B}, where we check that our scheme is valid even for
the resonance state.  As the third model, in Sec.~\ref{sec:3C} we
consider a two-channels problem to generate a bound state in the
higher channel which decays into the lower channel.  Finally we employ
a model of an unstable ``bound state'' composed of an unstable
particle and a stable particle in Sec.~\ref{sec:3D} and show the
properties of the ``bound state'' in terms of the wave function and
compositeness extracted from the scattering amplitude.

\subsection{Bound states in a single-channel case}
\label{sec:3A}

In this subsection we consider bound states in a single-channel case.
The masses of two particles are fixed as $m = 1115.7 \mev$ and $M = 35
m$, and the interaction is taken to be a local one:
\begin{equation}
  V ( E; \, r ) = \frac{v ( E )}{1 + \exp [ ( r - R ) / a ]} ,
  \label{eq:VA}
\end{equation}
where $a$ and $R$ are parameters to fix the range of the interaction
and $v ( E )$ controls the strength of the interaction.  In this study
we fix the parameters as $a = 0.5 \fm$ and $R = 3.6 \fm$, and employ
the following expression of the energy dependent part as
\begin{equation}
  v ( E ) = v_{0} + v_{1} ( E - E_{0} ) .
  \label{eq:vE_A}
\end{equation}
with constants $v_{0}$ and $v_{1}$ and a certain energy scale $E_{0}$
to be determined later.  In this study we fix $v_{0} = -35 \mev$ and
$v_{1}$ is allowed to shift in a certain range so as to produce the
energy dependence of the interaction.  We note that in this
construction the difference between the nonrelativistic and
semirelativistic cases are tiny.  In this subsection we only consider
the nonrelativistic case.

\begin{figure}[!t]
  \centering
  \Psfig{8.6cm}{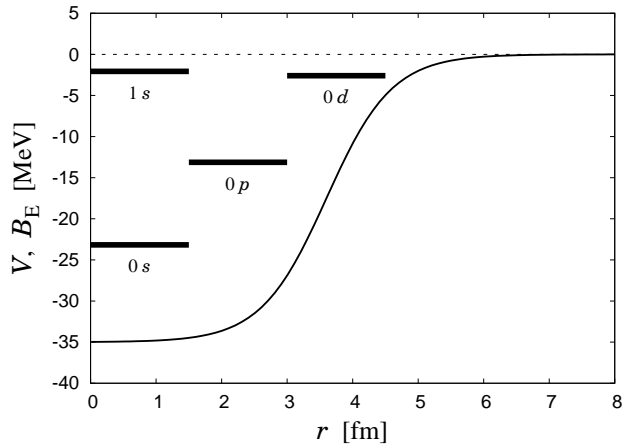} 
  \caption{Interaction~\eqref{eq:VA} as a function of the radial
    coordinate $r$ with its strength $v_{0} = -35 \mev$ and $v_{1} =
    0$.  We also show the binding energies of the bound states $B_{\rm
      E} \equiv m + M - E_{\rm pole}$: $B_{\rm E} ( 0 s ) = 23.2
    \mev$, $B_{\rm E} ( 0 p ) = 13.1 \mev$, $B_{\rm E} ( 0 d ) = 2.6
    \mev$, and $B_{\rm E} ( 1 s ) = 2.1 \mev$.}
  \label{fig:A_int}
\end{figure}

\begin{figure}[!b]
  \centering
  \Psfig{8.6cm}{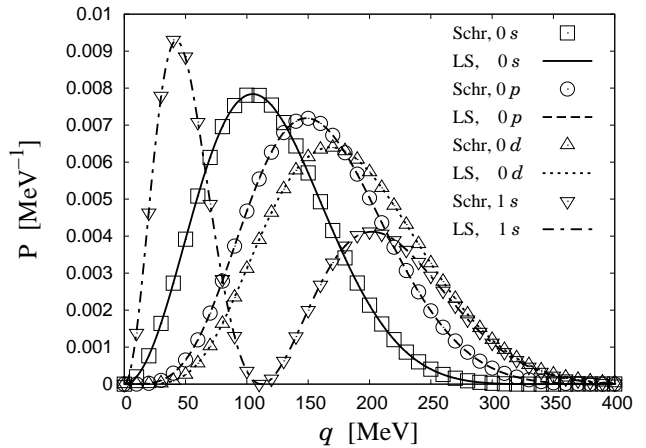} 
  \caption{Density distributions $\Rho ( q )$ obtained from the \Schr
    equation (Schr) and from the Lippmann--Schwinger equation (LS)
    with the interaction strength $v_{0} = -35 \mev$ and $v_{1} = 0$.
    The solutions of the \Schr equation are normalized by hand, while
    those of the Lippmann--Schwinger equation are automatically
    scaled. }
  \label{fig:A_WF}
\end{figure}

We first fix $v_{1} = 0$ and solve the \Schr
equation~\eqref{eq:Schr_final} and Lippmann--Schwinger
equation~\eqref{eq:LS_final}.  The interaction $V (r)$ is plotted as a
function of the radial coordinate $r$ in Fig.~\ref{fig:A_int}.  With
this interaction, we obtain four bound states $0 s$, $0 p$, $0 d$, and
$1 s$ as discrete eigenstates, whose binding energies, $B_{\rm E}
\equiv m + M - E_{\rm pole}$, are $B_{E} ( 0 s ) = 23.2 \mev$, $B_{\rm
  E} ( 0 p ) = 13.1 \mev$, $B_{\rm E} ( 0 d ) = 2.6 \mev$, and $B_{\rm
  E} ( 1 s ) = 2.1 \mev$.  These levels are shown in
Fig.~\ref{fig:A_int} as well.  For these four states, we compare the
wave functions from the \Schr equation in a usual manner and from the
Lippmann--Schwinger equation in our scheme developed in
Sec.~\ref{sec:2C}.  The density distributions from the wave functions
are shown in Fig.~\ref{fig:A_WF}.  Here we note that, while the wave
functions from the \Schr equation are normalized by hand so that their
compositeness is exactly unity, those from the Lippmann--Schwinger
equation are automatically scaled when solving the equation.  As one
can see from Fig.~\ref{fig:A_WF}, the wave function from the
Lippmann--Schwinger equation coincides with the normalized one from
the \Schr equation in all values of $q$ for each bound state.
Therefore, the compositeness $X$ evaluated from the residue of
scattering amplitude is automatically unity for every bound state.
Here we mention that the normalization of the wave function from the
residue of the scattering amplitude was already discussed in
Ref.~\cite{Hernandez:1984zzb}, in which it was proved that an
energy-independent interaction exactly gives $X = 1$.

\begin{figure}[!t]
  \centering
  \Psfig{8.6cm}{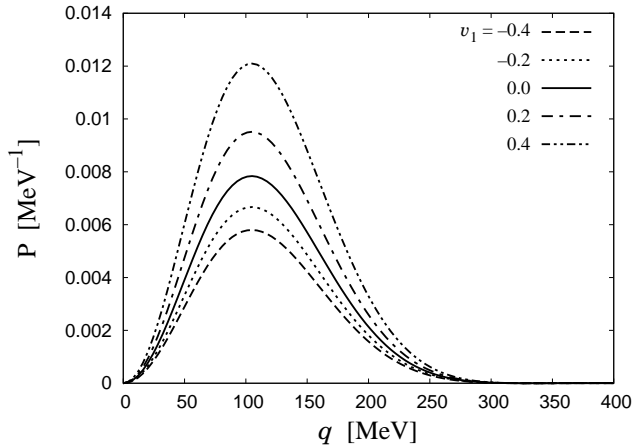} 
  \caption{Density distributions $\Rho ( q )$ 
    obtained from the Lippmann--Schwinger equation (LS) with several
    values of $v_{1}$.}
  \label{fig:A_WF_Edep}
\end{figure}

Next we change the value of $v_{1}$ and observe the response of the
wave function.  In the calculation of each bound state, we take $E_{0}
= E_{\rm pole}$ so that the eigenenergy does not change [see
  Eq.~\eqref{eq:vE_A}].  In Fig.~\ref{fig:A_WF_Edep} we show the
density distributions $\Rho ( q )$ obtained from the
Lippmann--Schwinger equation for energy-dependent interactions with
several values of $v_{1}$.  From the figure, we find that, although
the shape is the same for the density distributions with various
values of $v_{1}$, their peak heights become larger as $v_{1}$
increases.  Since the compositeness of the wave function with $v_{1} =
0$ is unity, when the interaction depends on the energy, the
compositeness from the scattering amplitude, which is automatically
scaled, deviates from unity.  From the figure, we can see that the
compositeness from the scattering amplitude becomes more (less) than
unity for $v_{1} > 0$ ($v_{1} < 0$).

We can interpret the behavior that the compositeness is smaller than
unity for $v_{1} < 0$ as the effect of the missing channel
contributions.  In order to see this, we here consider a one-body bare
state as the missing channel for simplicity; an extension to more than
one-body systems will be similar.  On the one hand, if there exists a
missing channel, its contribution $Z$ in Eq.~\eqref{eq:missing-I}
should be positive, $Z > 0$, and hence we should have $X < 1$
according to Eq.~\eqref{eq:sum_rule}.  On the other hand, since the
practical model space is a single two-body state only, this missing
channel should be implemented into the interaction, which inevitably
introduces the energy dependence to the interaction (see
Ref.~\cite{Sekihara:2014kya}).  Actually, for the one-body bare state,
its energy dependence is of the form of
\begin{equation}
  \tilde{V}_{\rm bare} ( E ; \, \bm{q}^{\prime} , \, \bm{q} )
  = \frac{g_{0}^{2}}{E - M_{0}} ,
\end{equation}
in momentum space, with a real coupling constant $g_{0}$ and the mass
$M_{0}$ for the one-body bare state.  This $\tilde{V}_{\rm bare}$ is
added to the usual energy-independent interaction.  As a result, since
we have $d \tilde{V}_{\rm bare} / d E ( E_{\rm pole} ) < 0$ regardless
of the values of $E_{\rm pole}$, $g_{0}$, and $M_{0}$, we have $d v /
d E ( E_{\rm pole} ) < 0$ for the interaction into which the missing
channel is implemented.  This explains why we obtain compositeness
smaller than unity, $X < 1$, for $v_{1} < 0$.

\begin{figure}[!t]
  \centering
  \Psfig{8.6cm}{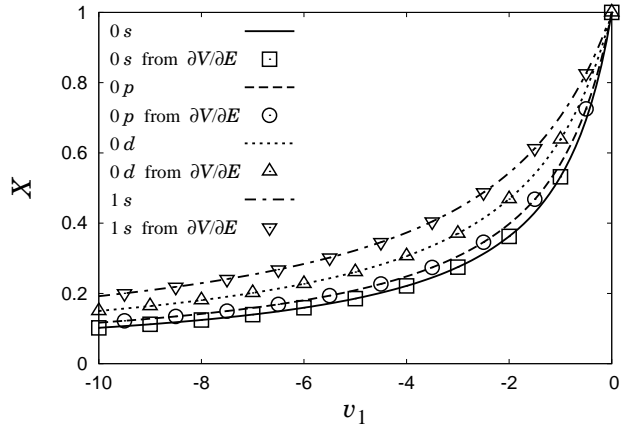} 
  \caption{Compositeness $X$ as a function of $v_{1}$.  Lines and
    points are obtained from the formulae in Eqs.~\eqref{eq:X_gamma}
    and \eqref{eq:X_E-A}, respectively, with the wave function from
    the scattering amplitude.}
  \label{fig:A_norm}
\end{figure}

We can discuss the relation between the compositeness from the
scattering amplitude and the energy-dependent interaction in a
different way by evaluating the compositeness $X$ as a function of
$v_{1}$.  First, in Fig.~\ref{fig:A_norm} we show the behavior for the
compositeness from the scattering amplitude as lines.  The
compositeness $X$ decreases when $v_{1}$ takes negatively larger
values, which can be interpreted as a larger missing-channel
contribution.  Next, we calculate the compositeness by the method
developed for an energy-dependent interaction.  Actually, according to
a discussion on an energy-dependent interaction in, {\it e.g.},
Refs.~\cite{Formanek:2003, Miyahara:2015bya}, the norm for the total
wave function, as a integral of the density with respect to the whole
coordinate space, should be modified as~\cite{Miyahara:2015bya}
\begin{equation}
  N = \int d^{3} r \psi ^{\ast} ( \bm{r} ) \left [ 1 -
    \frac{\prt V}{\prt E} ( E_{\rm pole} ; \, r ) \right ]
  \psi ( \bm{r} ) ,
  \label{eq:N_from_continuity}
\end{equation}
where $\psi ( \bm{r} )$ is the two-body wave function.  In the
right-hand side of the above equation, the first term is the norm of
the two-body wave function, which is nothing but the compositeness.
We express this first term as $X_{\prt V / \prt E}$.  The second term
containing the derivative $\prt V / \prt E$ is an additional term so
that the continuity equation from the wave function can be
hold.\footnote{The appearance of the derivative $\prt V / \prt E$ in
  the sum rule~\eqref{eq:N_from_continuity} was pointed out also in
  Ref.~\cite{Sekihara:2010uz} in terms of the conservation of a
  quantum number in the system.}  Since the present model space is
only a single two-body channel, the second term can be interpreted as
the missing channel contribution, which is not explicit degrees of
freedom.  Then, taking the norm for the total wave function as $N =
1$, we can calculate $X_{\prt V / \prt E}$ as
\begin{equation}
  X_{\prt V / \prt E} = 1 + \int d^{3} r \psi ^{\ast} ( \bm{r} ) 
    \frac{\prt V}{\prt E} ( E_{\rm pole} ; \, r ) 
  \psi ( \bm{r} ) .
\end{equation}
In terms of the wave function in momentum space, we can rewrite this
as
\begin{align}
  & X_{\prt V / \prt E}
  \notag \\
  & = 1 + \int \frac{d^{3} q}{( 2 \pi )^{3}}
  R ( q ) Y_{L M}^{\ast} ( \hat{q} ) \int \frac{d^{3} q^{\prime}}{( 2 \pi )^{3}}
  R ( q^{\prime} ) Y_{L M} ( \hat{q}^{\prime} ) 
  \notag \\
  & \phantom{= 1 + } \times
  \sum _{L^{\prime} = 0}^{\infty} \frac{\prt V_{L^{\prime}}}{\prt E}
  ( E_{\rm pole} ; \, q, \, q^{\prime} ) ( 2 L^{\prime} + 1 )
  P_{L^{\prime}} ( \hat{q}^{\prime} \cdot \hat{q} )
  \notag \\
  & = 1 + \int _{0}^{\infty} d q \frac{q^{2}}{2 \pi ^{2}} R ( q )
  \int _{0}^{\infty} d q^{\prime} \frac{q^{\prime \, 2}}{2 \pi ^{2}}
  R ( q^{\prime} )
  \notag \\
  & \phantom{= 1 +} \times
  \frac{\prt V_{L}}{\prt E}
  ( E_{\rm pole} ; \, q, \, q^{\prime} ) ,
  \label{eq:X_E-A}
\end{align}
where we have performed the integral with respect to the solid angles
by using the relations in Eqs.~\eqref{eq:Y_norm} and \eqref{eq:PYY}.
We evaluate $X_{\prt V / \prt E}$ by using the wave function from the
scattering amplitude, and show in Fig.~\ref{fig:A_norm} the behavior
for the compositeness from the formula~\eqref{eq:X_E-A} as points.  As
one can see, the points exactly lies on the line for each bound state.
Therefore, the two-body wave function from the scattering amplitude
correctly takes into account the effect of the additional term
in~\eqref{eq:X_E-A}, which can be interpreted as the missing channel
contribution for the system.

Here we emphasize that we will not obtain such an automatically scaled
wave function when we solve the \Schr equation with an energy
dependent interaction and normalize the wave function \naively within
the explicit model space.  In this sense, solving the
Lippmann--Schwinger equation at the bound state pole is equivalent to
evaluating the two-body wave function of the bound state, where the
effect of an energy-dependent interaction is also taken into account.

The above results are obtained with the nonrelativistic form for the
energy of the two-body state~\eqref{eq:Ej_NR}.  Here we note that with
the semirelativistic form~\eqref{eq:Ej_SR} we obtain the almost same
results for the wave functions and compositeness compared to the
nonrelativistic case above.  In particular, the compositeness is unity
for $v_{1} = 0$ but becomes less than unity for $v_{1} < 0$ also in
the semirelativistic case.

In summary, we can extract the two-body wave function from the
scattering amplitude as the residue of the amplitude at the pole
position of the bound state.  The scattering amplitude is a solution
of the Lippmann--Schwinger equation, and the two-body wave function
from the amplitude is automatically scaled.  In particular, for the
two-body wave functions from the amplitude, the compositeness deviates
from unity when the interaction depends on the energy, which can be
interpreted as the missing-channel contributions.  The present results
indicate that we can elucidate the hadron structure in terms of the
hadronic molecules from the hadron--hadron scattering amplitude,
assuming that the energy dependence of the hadron--hadron interaction
originates from missing channels which are not taken into account as
explicit degrees of freedom.  However, almost all of the interesting
hadrons are unstable in strong interaction.  Therefore we have to
consider cases of resonance states in detail and have to clarify the
relation between the wave functions from the \Schr equation and from
the Lippmann--Schwinger equation, which is the subject in the
following subsections.

\subsection{A resonance state in a single-channel case}
\label{sec:3B}

Next we consider a resonance state in a single-channel problem.  We
fix the masses of two particles as $m = M = 938.9 \mev$, and construct
the interaction between them in the form:
\begin{equation}
  V ( r; \, E ) = v ( E ) \left ( 2 e^{- r^{2} / b_{1}^{2}} -
  e^{- r^{2} / b_{2}^{2}} \right ) ,
  \label{eq:VB}
\end{equation}
where $b_{1}$ and $b_{2}$ are parameters to fix the interaction range
and $v ( E )$ controls the strength of the interaction.  This
interaction has an attractive core and a penetration barrier if $0 <
b_{1} < b_{2}$ and $v ( E ) < 0$.  In this study we fix the
interaction range as $b_{1} = 2.5 \fm$ and $b_{2} = 5.0 \fm$, and
employ the interaction strength $v (E)$ in Eq.~\eqref{eq:vE_A}.  The
constants $v_{0}$ and $v_{1}$ and a certain energy scale $E_{0}$ in $v
(E)$ are determined later.  In this subsection we consider only the
nonrelativistic case.  The calculation for a resonance state is done
in the complex scaling method.  We take the angle of the complex
scaling as $\theta = 20^{\circ}$ unless explicitly mentioned, but the
resonance pole position and compositeness from the scattering
amplitude do not depend on $\theta$.

\begin{figure}[!t]
  \centering
  \Psfig{8.6cm}{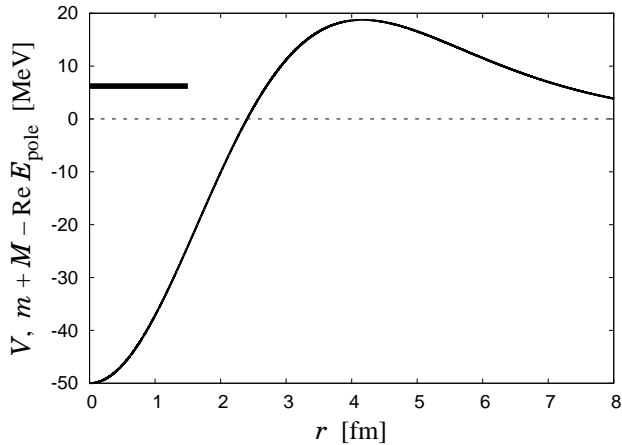} 
  \caption{Interaction $V$~\eqref{eq:VB} as a function of the radial
    coordinate $r$ with its parameters $v_{0} = - 50 \mev$, $v_{1} =
    0$, $b_{1} = 2.5 \fm$ and $b_{2} = 5.0 \fm$.  We also show the
    eigenenergy of the discrete resonance state measured from the
    threshold, $m + M - \text{Re} \, E_{\rm pole}$.}
  \label{fig:B_int}
\end{figure}

First we consider the case of an energy-independent interaction with
$v_{1} = 0$.  If the interaction strength $v_{0}$ is negatively
large enough, the interaction generates a stable bound state.
However, if $v_{0}$ is not so negatively large, this could generate an
unstable resonance state.  Actually, when we fix $v_{0} = - 50 \mev$,
we obtain a resonance state at the pole position $E_{\rm pole} =
1884.0 - 0.1 i \mev$ in the second Riemann sheet, $6.2 \mev$ above the
threshold $m + M = 1877.8 \mev$.  The behavior of the interaction in
coordinate space is shown in Fig.~\ref{fig:B_int} together with the
eigenenergy of the resonance state.

\begin{figure}[!t]
  \centering
  \Psfig{8.6cm}{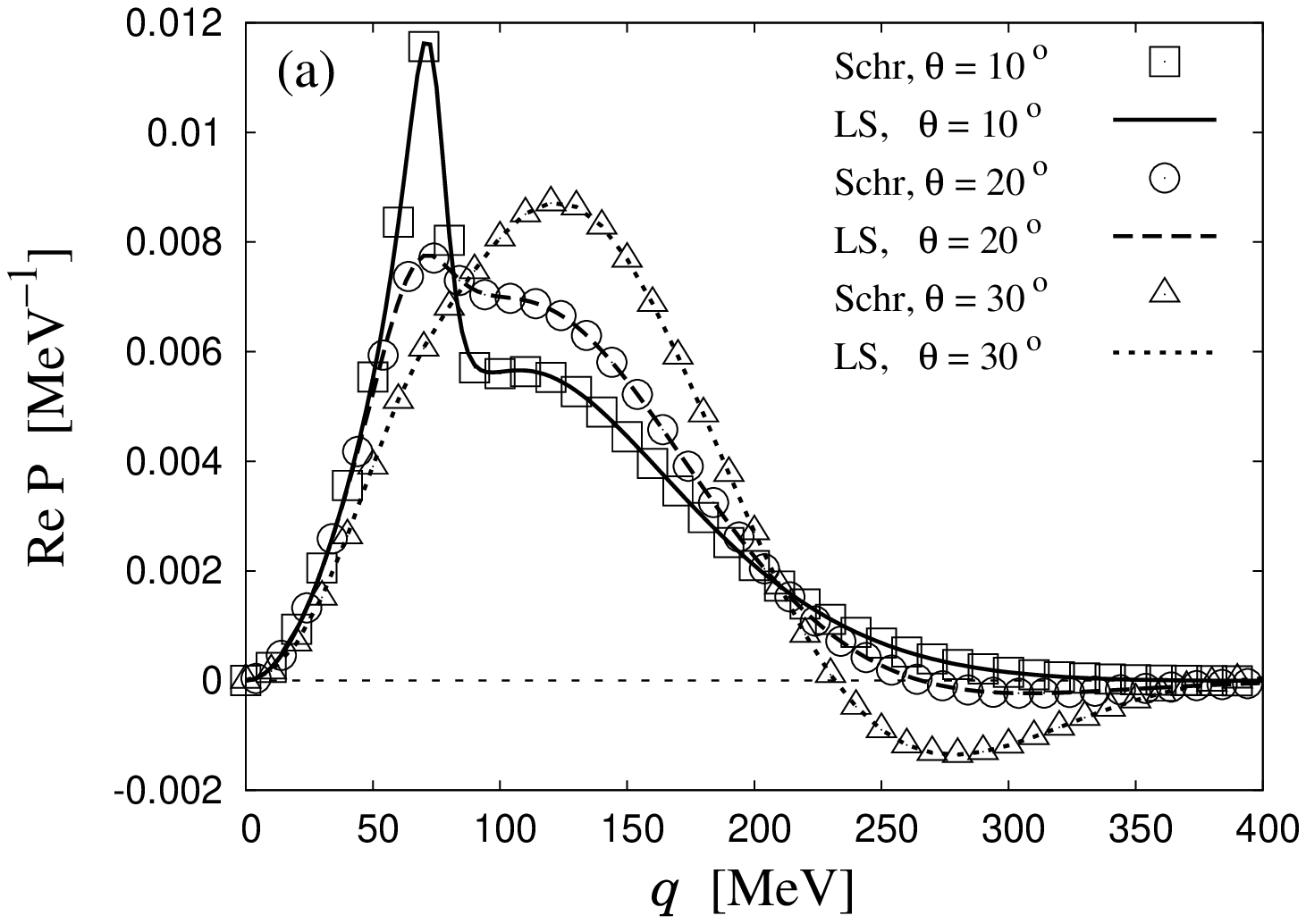} \\
  \Psfig{8.6cm}{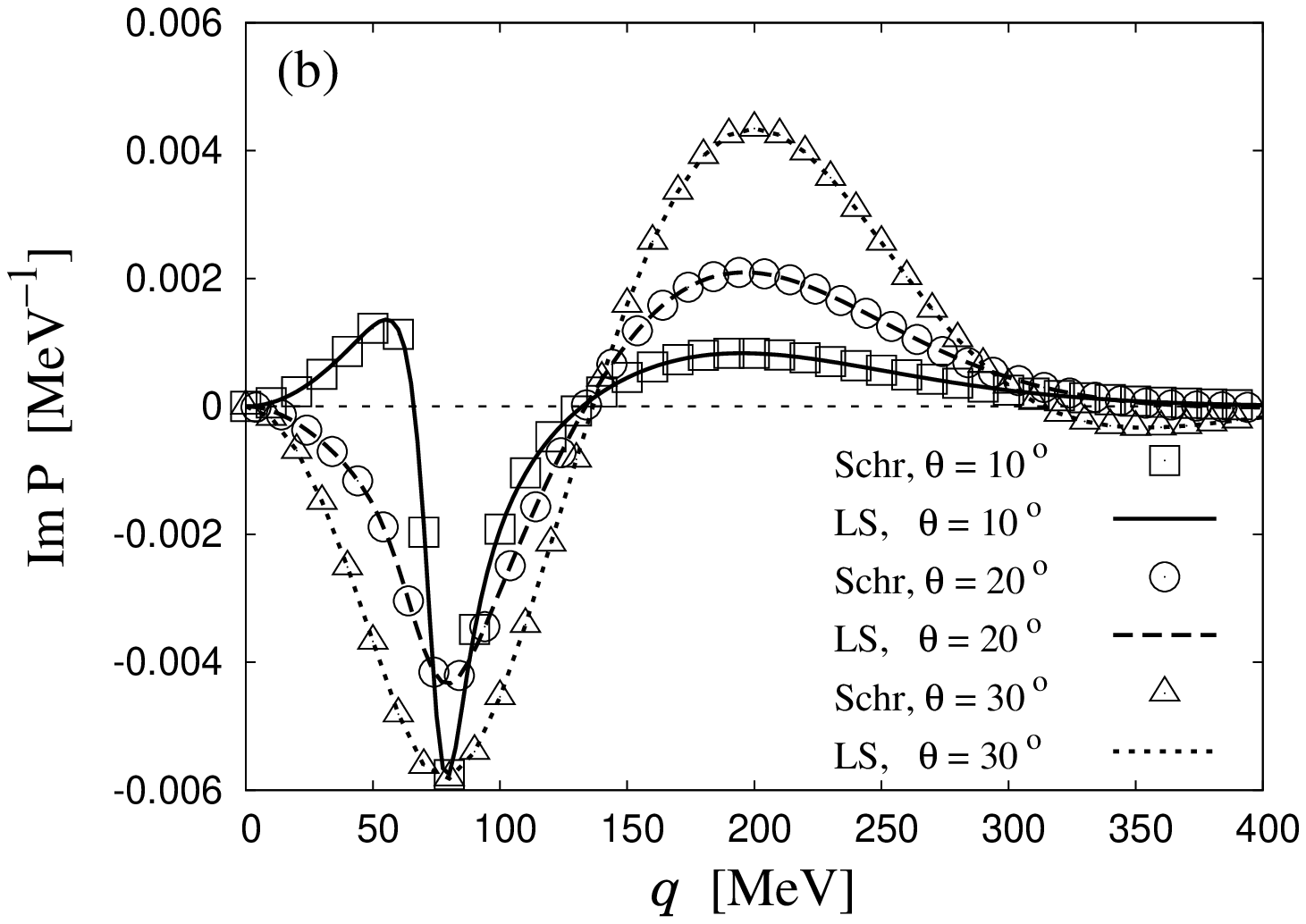} 
  \caption{(a) Real and (b) imaginary parts of the density
    distributions $\Rho ( q )$ obtained from the \Schr equation (Schr)
    and Lippmann--Schwinger equation (LS).  The scaling angle is
    $\theta = 10^{\circ}$, $20^{\circ}$, and $30^{\circ}$.}
  \label{fig:B_WF}
\end{figure}

For this resonance state, we calculate the wave function in two
approaches; one is solving the \Schr equation~\eqref{eq:Schr_CSM} with
its normalization by hand to be unity in terms of
Eqs.~\eqref{eq:Xj_CSM_Schr} and \eqref{eq:norm_CSM}, and the other one
is solving the Lippmann--Schwinger equation~\eqref{eq:LS_CSM} at the
resonance pole position, which cannot require any artificial scaling
of the wave function.  We numerically calculate the wave function in
both approaches with the complex-scaling angles $\theta = 10^{\circ}$,
$20^{\circ}$, and $30^{\circ}$, and show the density distribution
$\Rho ( q )$~\eqref{eq:X_CSM} for the resonance in
Fig.~\ref{fig:B_WF}.  Since the density distribution for the resonance
inevitably has an imaginary part, we plot both the real and imaginary
parts in Fig.~\ref{fig:B_WF}.\footnote{In the complex scaling method
  with finite $\theta$, the density distribution $\Rho ( q )$ becomes
  complex even for a stable bound state, since we have to calculate
  $\Rho ( q )$ with the complex-number-squared wave function in the
  complex scaling method.  However, for the stable bound state we can
  make the density distribution a real number by taking $\theta =
  0^{\circ}$, which cannot be taken for resonance states.}
Interestingly, the density distributions for the resonance in two
approaches coincide with each other for every value of the angle
$\theta$.  This means that we can obtain the correctly normalized
two-body wave function from the scattering amplitude even for a
resonance state, as discussed in Ref.~\cite{Hernandez:1984zzb}.

Here we note that, in general, the density distribution as well as the
wave function depends on the angle of the complex scaling $\theta$, as
shown in Fig.~\ref{fig:B_WF}.  In particular, for smaller $\theta$ the
density distribution shows a peak structure at $q \sim 75 \mev$.  This
can be understood with the expression in Eq.~\eqref{eq:X_CSM}.
Namely, for smaller $\theta$, the denominator of the density
distribution, $[E_{\rm pole} - \mathcal{E} ( q e^{- i \theta} )]^{2}$,
becomes almost zero at $q \sim 75 \mev$, which is nothing but the
relative momentum of two daughter particles from the decay of the
resonance.  This almost-zero denominator $[E_{\rm pole} - \mathcal{E}
  ( q e^{- i \theta} )]^{2}$ brings the peak structure in the density
distribution.

We emphasize that, although the wave function depends on the angle
$\theta$, the compositeness as the integral of the density
distribution does not depend on $\theta$.  In the present case, the
compositeness is exactly unity for every value of $\theta$.  We can
explain this fact by interpreting the complex-scaled momentum $q e^{-
  i \theta}$ as the change of the contour for the momentum integral of
the range $[ 0, \, \infty )$ to the rotated one with the angle $-
\theta$.  In this sense, when we calculate a matrix element of a
certain operator with the resonance wave function in the complex
scaling method, the matrix element does not depend on the angle
$\theta$ as long as the operator is properly complex scaled.

\begin{figure}[!t]
  \centering
  \Psfig{8.6cm}{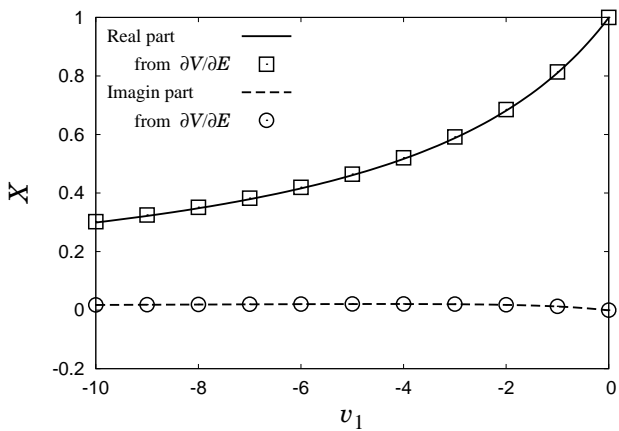} 
  \caption{Compositeness $X$ as a function of $v_{1}$.  Lines and
    points are obtained from the formulae in Eqs.~\eqref{eq:X_CSM} and
    \eqref{eq:X_E-B}, respectively, with the wave function from the
    scattering amplitude.}
  \label{fig:B_norm}
\end{figure}

Finally we introduce the energy dependence to the interaction.  As in
the previous subsection, we take $E_{0} = E_{\rm pole}$ so that the
pole position does not change.  Then, we expect that the compositeness
from the scattering amplitude will deviate from unity, as discussed in
the previous subsection.  Actually, as shown in Fig.~\ref{fig:B_norm},
where we plot the compositeness from the scattering
amplitude~\eqref{eq:X_CSM} for $v_{1} < 0$ as lines, the compositeness
behaves similarly to the case of stable bound states in
Fig.~\ref{fig:A_norm}.  The behavior implies that missing states
rather than the explicit two-body state compose the resonance state
more dominantly for larger $v_{1}$.  We also note that, since the
present state is an unstable resonance, the compositeness in
Fig.~\ref{fig:B_norm} has an imaginary part, although the imaginary
part is negligible compared to the real part.

We then compare the compositeness in Eq.~\eqref{eq:X_CSM} in
Fig.~\ref{fig:B_norm} with that from the continuity equation in the
energy-dependent interaction~\cite{Formanek:2003, Miyahara:2015bya}.
The extension of Eq.~\eqref{eq:N_from_continuity} to a resonance state
has been done in Ref.~\cite{Miyahara:2015bya}, and the compositeness
from the continuity equation, $X_{\prt V / \prt E}$, for the resonance
becomes
\begin{align}
  & X_{\prt V / \prt E}
  \notag \\
  & = 1 + e^{- 6 i \theta} \int _{0}^{\infty} d q \frac{q^{2}}{2 \pi ^{2}}
  R ( q e^{- i \theta} )
  \int _{0}^{\infty} d q^{\prime} \frac{q^{\prime \, 2}}{2 \pi ^{2}}
  R ( q^{\prime} e^{- i \theta} )
  \notag \\
  & \phantom{= 1 +} \times
  \frac{\prt V_{L}}{\prt E}
  ( E_{\rm pole} ; \, q e^{- i \theta}, \, q^{\prime} e^{- i \theta}) ,  
  \label{eq:X_E-B}
\end{align}
where we have performed the complex scaling.  The result of $X_{\prt V
  / \prt E}$ with the wave function from the scattering amplitude is
shown in Fig.~\ref{fig:B_norm} as points.  As one can see, the points
exactly lies on the lines both in the real and imaginary parts.
Therefore, the present result indeed indicates that the wave function
from the scattering amplitude correctly takes into account the effect
of the energy-dependent interaction as the additional term containing
the derivative in~\eqref{eq:X_E-B}.

In summary, even for a resonance state in a single-channel problem,
its automatically scaled two-body wave function is obtained from the
residue of the scattering amplitude at the pole position.  Although
the wave function itself depends on the scaling angle $\theta$ in the
complex scaling method, the compositeness as the integral of the wave
function squared does not depend on $\theta$.  The compositeness for
the resonance state from the scattering amplitude deviates from unity
when we take into account the energy dependence of the interaction,
which can be interpreted as the implementation of missing-channel
contributions, as in the case of stable bound states.

\subsection{A resonance state in a coupled-channels case}
\label{sec:3C}

In this subsection we extend our discussion to a resonance state in a
two-channels case.  The masses of the system are fixed as $m_{1} =
495.7 \mev$, $M_{1} = 938.9 \mev$, $m_{2} = 138.0 \mev$, and $M_{2} =
1193.1 \mev$.  The interaction is fixed in the form:
\begin{equation}
  V_{j k} ( r ; \, E ) = v ( E ) C_{j k}
  e ^{ - r^{2} / b^{2} } ,
  \label{eq:VC}
\end{equation}
where $b$ is the range parameter of the interaction and $v ( E )$ and
$C_{j k}$ respectively control the strength of the interaction and
transition between different channels.  The expression of $v ( E )$ is
given in Eq.~\eqref{eq:vE_A} and $C_{j k}$ is
\begin{equation}
  C_{j k} =
  \left ( 
  \begin{array}{@{\,}cc@{\,}}
    1 & x \\
    x & 0
  \end{array}
  \right ) ,
\end{equation}
with a parameter $x$.  In this study we fix $b = 0.5 \fm$ and $v_{0} =
- 650 \mev$ in $v ( E )$, while $x$ is allowed to shift in a certain
range.  Throughout this subsection we take the angle of the complex
scaling as $\theta = 20^{\circ}$ in treating the resonance state.  In
this subsection we consider only the semirelativistic case; we have
checked that the nonrelativistic case gives similar behavior for the
wave function from the Lippmann--Schwinger equation.

\begin{figure}[!t]
  \centering
  \Psfig{8.6cm}{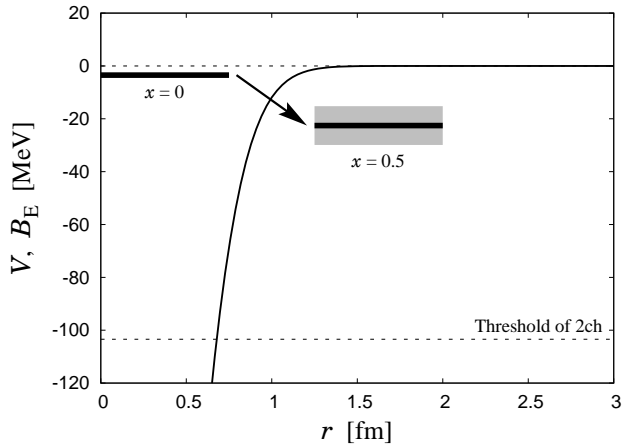} 
  \caption{Interaction $V_{11}$~\eqref{eq:VC} as a function of the
    radial coordinate $r$ with its strength $v_{0} = -650 \mev$ and
    $v_{1} = 0$.  We also show the binding energies of the discrete
    bound states $B_{\rm E} \equiv m_{1} + M_{1} - \text{Re} \, E_{\rm
      pole}$ with $x = 0$ and $x = 0.5$.  The shaded band for $x =
    0.5$ indicates the range of $B_{\rm E} \pm \text{Im} \, E_{\rm
      pole}$.}
  \label{fig:C_int}
\end{figure}

We first fix $v_{1} = 0$ and $x = 0$, and solve the
Lippmann--Schwinger equation~\eqref{eq:LS_final} by taking into
account only the channel $1$.  The interaction $V_{11} ( r )$ is
plotted as a function of the radial coordinate $r$ in
Fig.~\ref{fig:C_int}.  In this condition we have a $0 s$ bound state
at the eigenvalue $E_{\rm pole} = 1431.1 \mev$ with the binding energy
$B_{\rm E} \equiv m_{1} + M_{1} - E_{\rm pole} = 3.5 \mev$.  For this
bound state, we have checked that we can extract the two-body wave
function as the residue of the scattering amplitude at the pole of the
bound state, with the compositeness exactly unity.

\begin{table}[!b]
  \caption{Properties of the resonance state in the coupled-channels
    interaction~\eqref{eq:VC}.  The parameters are fixed as $b = 0.5
    \fm$, $v_{0} = -650 \mev$, $v_{1} = 0$, $x = 0.5$, and $\theta =
    20^{\circ}$.  The binding energy $B_{\rm E}$ and width $\Gamma$
    are defined as $B_{\rm E} \equiv m_{1} + M_{1} - \text{Re} \,
    E_{\rm pole}$ and $\Gamma \equiv - 2 \, \text{Im} \, E_{\rm
      pole}$, respectively. }
  \label{tab:1}
  \begin{ruledtabular}
    \begin{tabular*}{5.6cm}{@{\extracolsep{\fill}}lc}
      $B_{\rm E}$ [MeV] & $22.6$ 
      \\
      $\Gamma$ [MeV] & $14.7$
      \\
      $X_{1}$ & $0.99 - 0.08 i$
      \\
      $X_{2}$ & $0.01 + 0.08 i$
      \\
      $X_{1} + X_{2}$ & $1.00 + 0.00 i$
      \\
      $U$ & $0.07$
      \\
      $\tilde{X}_{1}$ & $0.93$
      \\
      $\tilde{X}_{2}$ & $0.07$
      \\
    \end{tabular*}
  \end{ruledtabular}
\end{table}

\begin{figure}[!t]
  \centering
  \Psfig{8.6cm}{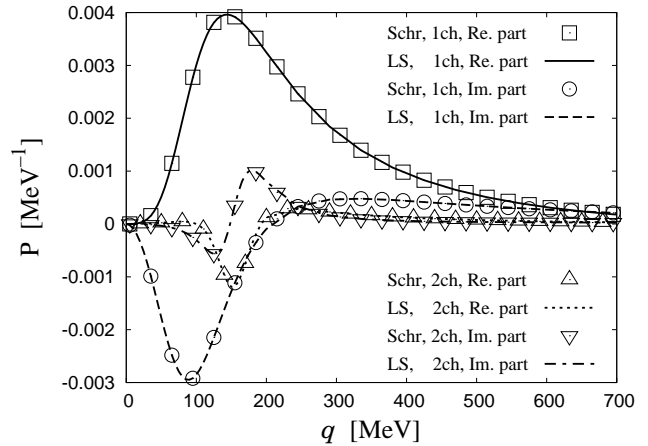} 
  \caption{Density distributions $\Rho ( q )$ obtained from the \Schr
    equation (Schr) and Lippmann--Schwinger equation (LS) with the
    interaction parameters $v_{0} = -650 \mev$, $v_{1} = 0$ and $x =
    0.5$.  The scaling angle is $\theta = 20^{\circ}$.}
  \label{fig:C_WF}
\end{figure}

Now let us switch on the coupling with nonzero $x$.  We here take $x =
0.5$, with which the $0 s$ bound state in the channel $1$ becomes a
resonance state.  The eigenenergy $E_{\rm pole}$ is $1412.0 - 7.3
\mev$, which is also shown in Fig.~\ref{fig:C_int}.  The properties of
this resonance state is listed in Table~\ref{tab:1}.  In
Fig.~\ref{fig:C_WF}, we show the density distributions from the \Schr
equation~\eqref{eq:Schr_CSM} and Lippmann--Schwinger
equation~\eqref{eq:LS_CSM}.  We note that, while the wave function
from the \Schr equation is normalized by hand so that the sum of the
compositeness in two channels is exactly unity, that from the
Lippmann--Schwinger equation is automatically scaled.  Amazingly, the
wave functions from two equations coincide with each other.  In
particular, since the wave function from the \Schr equation is
normalized, we can see that the wave function from the
Lippmann--Schwinger equation is also correctly normalized.  This
result indicates that even for a resonance state in a coupled-channels
problem we can extract the wave function from the scattering amplitude
at the pole position.  In Table~\ref{tab:1}, we also list the value of
the compositeness from the Lippmann--Schwinger equation.  From the
result, the compositeness in the channel $1$, $X_{1}$, dominates the
normalization $X_{1} + X_{2} = 1$, which implies that this resonance
state is a bound state of two particles in the channel $1$ with a
coupling to the decaying channel $2$.  We have checked that the
normalization $X_{1} + X_{2} = 1$ is obtained for resonance wave
functions from the scattering amplitude with any different value of
the parameter $x$.

In order to interpret the complex compositeness of each channel $X_{1,
  2}$, we calculate $\tilde{X}_{1, 2}$ together with $U$ in
Eqs.~\eqref{eq:XtildeZtilde} and \eqref{eq:XtildeU}.  The values of
these quantities for the present resonance state are listed in
Table~\ref{tab:1}.  Because the present state gives $U = 0.07 \ll 1$,
according to the discussion given after Eq.~\eqref{eq:Xtilde_sum} we
can safely interpret $\tilde{X}_{1}$ ($\tilde{X}_{2}$) as the
probability to find the two-body component of channel $1$ ($2$).  From
the numerical result, we can conclude that this resonance is indeed a
bound state of two particles in channel $1$.

\begin{figure}[!t]
  \centering
  \Psfig{8.6cm}{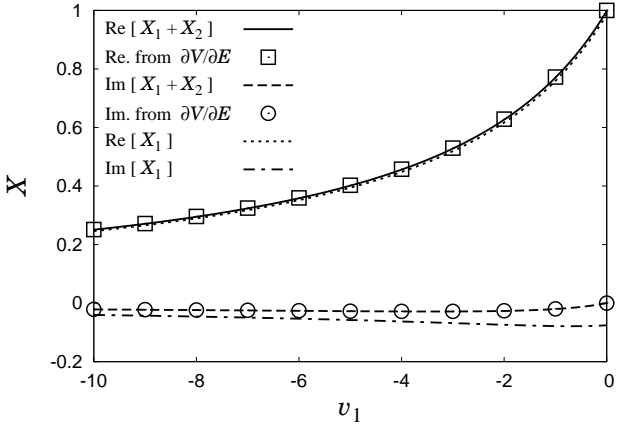} 
  \caption{Compositeness $X$ as a function of $v_{1}$.  Lines and
    points are obtained from the formulae in Eqs.~\eqref{eq:X_CSM} and
    \eqref{eq:X_E-C}, respectively, with the wave function from the
    scattering amplitude.}
  \label{fig:C_norm}
\end{figure}

Next we change the value of $v_{1}$ so as to introduce the energy
dependence of the interaction.  We take $E_{0} = E_{\rm pole}$ so that
the pole position does not change.  The wave function is calculated
from the scattering amplitude.  The behavior of the compositeness in
channel $1$ and the sum $X_{1} + X_{2}$ are shown in
Fig.~\ref{fig:C_norm}.  As one can see, the real part of $X_{1} +
X_{2}$ as well as that of $X_{1}$ decreases when $v_{1}$ takes
negatively larger values, which implies that the contribution form
missing states becomes important in this condition.  This result is
consistent with that in the previous subsections.  In addition, we
note that, while sum $X_{1} + X_{2}$ becomes unity for $v_{1} = 0$, it
becomes complex for $v_{1} \ne 0$.

The value of the compositeness is compared with that from the
continuity equation in the energy-dependent interaction.  Actually,
for a resonance state in a coupled-channels problem, one can
straightforwardly extend the formula~\eqref{eq:X_E-B}, which result in
\begin{align}
  & X_{\prt V / \prt E}
  = 1 + e^{- 6 i \theta} \sum _{j , k}
  \int _{0}^{\infty} d q \frac{q^{2}}{2 \pi ^{2}}
  R_{j} ( q e^{- i \theta} )
  \notag \\
  & \times \int _{0}^{\infty} d q^{\prime} \frac{q^{\prime \, 2}}{2 \pi ^{2}}
  R_{k} ( q^{\prime} e^{- i \theta} )
  \frac{\prt V_{L, j k}}{\prt E}
  ( E_{\rm pole} ; \, q e^{- i \theta}, \, q^{\prime} e^{- i \theta}) .
  \label{eq:X_E-C}
\end{align}
The result of $X_{\prt V / \prt E}$ with the wave function from the
scattering amplitude is shown in Fig.~\ref{fig:C_norm} as points.
From the figure, the points exactly lies on the lines of $X_{1} +
X_{2}$, which means that the sum $X_{1} + X_{2}$ correctly reproduces
the effect of the energy-dependent interaction in~\eqref{eq:X_E-C}.

In summary, even for an unstable resonance state in a coupled-channels
case, we can extract its two-body wave function from the scattering
amplitude as the residue of the scattering amplitude at the resonance
pole.  The wave function from the scattering amplitude is
automatically scaled in calculating the Lippmann--Schwinger equation
at the resonance pole.  In particular, we find that the sum of the
compositeness is exactly unity for the energy-independent interaction,
but it deviates from unity for the energy-dependent interaction, which
can be interpreted as the missing-channel contribution.  This result
indicates that our scheme is valid even for an unstable resonance
state in a coupled-channels problem, which will be essential when we
investigate the hadronic molecular components for hadronic resonances
in a coupled-channels approach.

\subsection{A ``bound state'' of an unstable constituent}
\label{sec:3D}

Finally we consider a ``bound state'' which contains an unstable
constituent.  This bound state is intrinsically unstable due to the
decay of the unstable constituent particle.  An example is a bound
state of the $\sigma$ meson and nucleon, if it existed, since the
$\sigma$ meson decays into $\pi \pi$ in the $\sigma N$ bound state as
well as in free space.  We evaluate the scattering amplitude of the
unstable particle $A$ and stable particle $B$ by introducing the
self-energy for $A$ in a method developed in
Ref.~\cite{Kamano:2008gr}.  Then, we extract the two-body wave
function of the $A B$ bound state from the scattering amplitude.  In
this subsection we take the semirelativistic formulation.

In order to calculate the two-body wave function of the $A B$ bound
state, we first describe the unstable constituent $A$.  In this study
we consider a case that a bare particle of mass $m_{\rm bare}$ couples
to a two-body decay channel in $s$ wave to be a physical $A$.  We
assume that the two particles in the decay channel have the same mass
$m_{d}$, which should satisfy $2 m_{d} < m_{\rm bare}$ so that $A$
decays.

\begin{figure}[t]
  \centering
  \PsfigII{0.18}{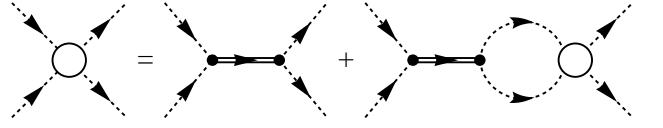} 
  \caption{Diagram for the description of the physical unstable
    particle $A$.  Double and dashed lines represent the bare particle
    and decay channel for $A$, respectively.}
  \label{fig:system_A}
\end{figure}

Suppose that the coupling of the bare particle for $A$ and the decay
channel is controlled by
\begin{equation}
  f ( q ) = \frac{\alpha \lambda ^{2}}{q^{2} + \lambda ^{2}}
\end{equation}
with the magnitude of the relative momentum in the decay channel $q$,
a coupling constant $\alpha$, and a cutoff $\lambda$.  Because of this
coupling, the bare particle becomes an unstable physical state $A$ in
the two-body system of the decay channel.  Actually, the physical
state $A$ appears as the pole of the scattering amplitude for the
decay channel, which can be described by the Lippmann--Schwinger
equation (see Fig.~\ref{fig:system_A}):
\begin{align}
  & \mathcal{T} ( E_{2} ; \, q^{\prime} , \, q )
  = \mathcal{V} ( E_{2} ; \, q^{\prime} , \, q )
  \notag \\
  & + \int \frac{d^{3} k}{( 2 \pi )^{3}}
  \frac{\mathcal{V} ( E_{2} ; \, q^{\prime} , \, k )
    \mathcal{T} ( E_{2} ; \, k , \, q )}{E_{2} - \mathcal{E}_{d} ( k )} ,
  \label{eq:Tbare}
\end{align}
with the total energy of the bare particle-decay channel system
$E_{2}$, the energy of the on-shell particles in the decay channel
$\mathcal{E}_{d} ( q ) \equiv 2 \sqrt{q^{2} + m_{d}^{2}}$, and the
``interaction'' between two particles in the decay channel via the
bare particle, $\mathcal{V}$:
\begin{equation}
  \mathcal{V} ( E_{2} ; \, q^{\prime} , \, q )
  \equiv \frac{f ( q^{\prime} ) f ( q )}{E_{2} - m_{\rm bare}} .
\end{equation}
After a simple algebra with ansatz $\mathcal{T} \propto f ( q^{\prime}
) f ( q )$, we can solve the Lippmann--Schwinger
equation~\eqref{eq:Tbare} and obtain the equation for the physical
mass of the unstable particle $A$, $m_{\rm phys}$, from its bare mass
$m_{\rm bare}$:
\begin{equation}
  m_{\rm phys} = m_{\rm bare} +
  \int \frac{d^{3} k}{( 2 \pi )^{3}}
  \frac{f ( k )^{2}}{m_{\rm phys} - \mathcal{E}_{d} ( k )} .
  \label{eq:mphys}
\end{equation}
We note that $m_{\rm phys}$ has an imaginary part when $2 m_{d} <
m_{\rm bare}$.

Let us fix the parameters $m_{\rm bare} = 600 \mev$, $m_{d} = 138.0
\mev$, $\alpha = 0.15 \mev ^{-1/2}$ and $\lambda = 600 \mev$.  In this
condition we obtain the physical mass~\eqref{eq:mphys} as $m_{\rm
  phys} = 422.7 - 52.0 i \mev$.  Interestingly, we can calculate the
compositeness of the decay channel for this physical unstable particle
$A$ from the scattering amplitude~\eqref{eq:Tbare} in our scheme,
which results in $X_{d} = 0.10 + 0.29 i$.  This indicates that we see
only subdominant component of the decay channel inside $A$ since its
absolute value, $| X_{d} | = 0.31$, is negligible compared to unity.

\begin{figure}[t]
  \centering
  \PsfigII{0.18}{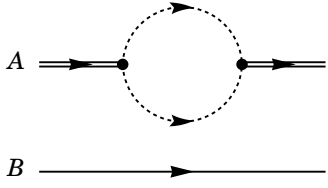} 
  \caption{Diagram for the self-energy of an unstable particle $A$ in
    the $A B$ propagator.  Dashed lines represent the decay channel
    for $A$.}
  \label{fig:system_AB}
\end{figure}

Now we consider the \Schr equation and Lippmann--Schwinger equation
for the two-body system of this unstable particle $A$ and the stable
particle $B$ of mass $M$.  Since the unstable particle $A$ decays and
its physical mass $m_{\rm phys}$ has an imaginary part, we should
rewrite both the equations in an appropriate way.  This can be done by
introducing the self-energy for the unstable particle $\Sigma ( E_{3}
; \, q )$, where $E_{3}$ is the eigenenergy of the whole system and
$q$ is the magnitude of the relative momenta between $A$ and $B$ (see
a diagram for the self-energy in Fig.~\ref{fig:system_AB}).  We employ
the approach developed in Ref.~\cite{Kamano:2008gr} and formulate the
self-energy as
\begin{align}
  \Sigma ( E_{3} ; \, q )
  = & \frac{m_{\rm bare}}{\sqrt{q^{2} + m_{\rm bare}^{2}}}
  \int \frac{d^{3} k}{( 2 \pi )^{3}}
  \frac{\mathcal{E}_{d} ( k )}{\sqrt{\mathcal{E}_{d}( k )^{2} + q^{2}}}
  \notag \\
  & \times
  \frac{f ( k )^{2}}{E_{3} - \sqrt{q^{2} + M^{2}}
    - \sqrt{ \mathcal{E}_{d}( k )^{2} + q^{2}}} .
\end{align}
By using this self-energy, we have to replace the two-body energy
$\mathcal{E} ( q )$ in the left-hand side of \Schr
equation~\eqref{eq:Schr_final} with
\begin{equation}
  \mathcal{E}_{\Sigma} ( E_{3} ; \, q )
  = \sqrt{q^{2} + m_{\rm bare}^{2}} + \Sigma ( E_{3} ; \, q )
  + \sqrt{q^{2} + M_{\phantom{j}}^{2}} .
  \label{eq:Eself}
\end{equation}
In a similar manner, the two-body energy $\mathcal{E} ( q )$ in the
Lippmann--Schwinger equation~\eqref{eq:LS_final} and in the
compositeness formulation~\eqref{eq:X_gamma} should be replaced with
the above $\mathcal{E}_{\Sigma} ( E_{3} ; \, q )$.  Then, momenta in
every equation are transformed into the complex-scaled values in the
complex scaling method so as to solve these equations for a resonance
state.

For the interaction between $A$ and $B$, we employ the Yukawa
function:
\begin{equation}
  V ( r ) = \beta \frac{e^{- \mu r}}{r} ,
\end{equation}
with the coupling constant $\beta$ and the interaction range $\mu$.
The Fourier transformation of this interaction is
\begin{equation}
  \tilde{V} ( q )
  = \int d^{3} r \, V ( r )
  e^{- i \bm{q} \cdot \bm{r}}
  = \frac{4 \pi \beta}{q^{2} + \mu ^{2}} ,
\end{equation}
However, in the semirelativistic case the Yukawa interaction leads to
an ultraviolet divergence for integrals.  In order to tame the
divergence, we introduce a form factor $\Lambda ^{2} / ( q^{2} +
\Lambda ^{2} )$ with a cutoff $\Lambda$ as
\begin{equation}
  \tilde{V} ( q )
  = \frac{4 \pi \beta}{q^{2} + \mu ^{2}}
  \frac{\Lambda ^{2}}{q^{2} + \Lambda ^{2}}
\end{equation}

Then, we consider a bound state of the unstable $A$ and stable $B$.
We fix the parameters as $M = 938.9 \mev$, $\beta = -2.0$, $\mu = 450
\mev$, and $\Lambda = 1.0 \gev$.  The parameters for the unstable $A$
are the same: $m_{\rm bare} = 600 \mev$, $m_{d} = 138.0 \mev$, $\alpha
= 0.15 \mev ^{-1/2}$ and $\lambda = 600 \mev$.  As a result, we obtain
an $s$-wave bound state of the unstable $A$ and stable $B$ at its pole
position $1363.8 - 32.2 i \mev$.

\begin{figure}[!t]
  \centering
  \Psfig{8.6cm}{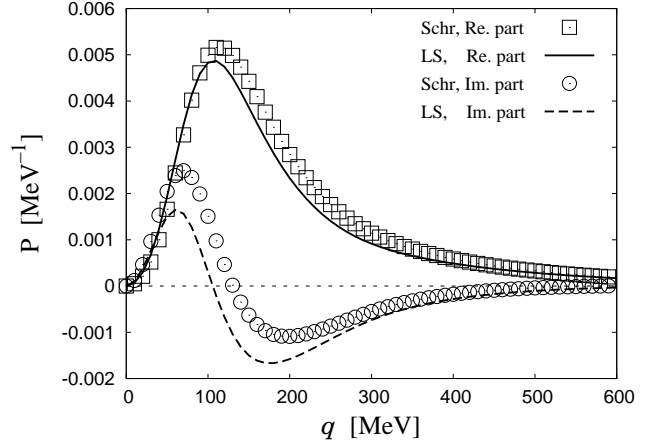} 
  \caption{Density distribution $\Rho ( q )$ obtained from the \Schr
    equation (Schr) and Lippmann--Schwinger equation (LS) for the
    bound state of unstable and stable particles.  The scaling angle
    is $\theta = 20^{\circ}$.}
  \label{fig:D_WF}
\end{figure}

We now solve the \Schr and Lippmann--Schwinger equations to obtain the
two-body wave function of the $A B$ bound state.  We show in
Fig.~\ref{fig:D_WF} the density distribution $\Rho ( q )$ calculated
from the wave function in two equations.  The one in the \Schr
equation is normalized to be unity by hand, while the one in the
Lippmann--Schwinger equation is extracted from the residue of the
scattering amplitude without scaling factor.  We note that the density
distribution as well as the wave function becomes complex since the
bound state is intrinsically resonance due to the unstable
constituent.

As one can see, the density distributions from two equations do not
coincide with each other.  Both the real and imaginary parts from the
Lippmann--Schwinger equation are smaller than those from the \Schr
equation.  Actually, the compositeness from the scattering amplitude
is evaluated as $X = 0.90 - 0.21 i$, which is unity for the wave
function from the \Schr equation.  This fact can be interpreted as the
effect of missing-channel contributions, in this case the decay
channel of the unstable constituent $A$.  In the present condition,
this missing-channel contribution is implemented as the energy
dependence of the self-energy $\Sigma$ for the unstable $A$ rather
than the interaction.

Here we note that the compositeness $X = 0.90 - 0.21 i$ is slightly
different from the value which satisfies the sum rule with $X_{d}$,
i.e., we have $X + X_{d} = 1.00 + 0.08 i \ne 1$.  This is because, in
the present formulation, if $A$ is inside the bound state, the field
renormalization constant for $A$ may change from the value in free
space, which is nothing but $1 - X_{d}$.  This is caused by the fact
that the self-energy for the particle $A$ inside the bound state
depends on the whole energy $E_{3}$ and the relative momentum between
$A$ and $B$, $q$.  Indeed, in free space the self-energy for $A$ is
$\Sigma ( m_{\rm phys} + M ; \, q = 0 )$, but this becomes $\Sigma (
E_{\rm pole} ; \, q )$ inside the bound state, which can change the
field renormalization constant for $A$ inside the bound state.
Actually, we find that, when we neglect $q$ dependence of the
self-energy $\Sigma$ and fix parameters so that $E_{\rm pole} \approx
m_{\rm phys} + M$, the sum rule $X + X_{d} = 1$ returns to be
satisfied.

In summary, from the scattering amplitude we can extract the resonance
wave function for a bound state which contains an unstable
constituent.  We have observed that the compositeness deviates from
unity due to the decay channel for the unstable constituent as a
missing contribution, although the sum rule of the compositeness is
broken slightly by the shift of the field renormalization constant for
the unstable constituent from the value in free space.  This
discussion will help us investigate, {\it e.g.}, the $\sigma N$ and
$\rho N$ components inside the $N^{\ast}$ and $\Delta ^{\ast}$
resonances in a forthcoming paper~\cite{Sekihara:2016}.

\section{Summary}
\label{sec:4}

In this study we have established a way to evaluate the two-body wave
functions of bound states, both in the stable and decaying cases, from
the residue of the scattering amplitude at the pole position.  An
important finding is that the two-body wave functions of the bound
states are automatically scaled when evaluated from the scattering
amplitude.  In particular, while the compositeness, defined as the
norm of the two-body wave function, is unity for energy-independent
interactions, it deviates from unity for energy-dependent
interactions, which can be interpreted as missing-channel
contributions.  We have checked that our scheme works correctly by
considering bound states in a single-channel problem and resonances in
three cases: single-channel, coupled-channels, and unstable
constituent cases.

We emphasize that the compositeness $X_{j}$ is not observable and
hence in general a model dependent quantity.  However, we can uniquely
determine it from the scattering amplitude once we fix the model
space, form of the kinetic energy $\mathcal{E}_{j}(q)$, and
interaction.  We also note that, while the resonance wave function
depends on the scaling angle $\theta$ in the complex scaling method,
the compositeness, or in general quantities as the integral of the
wave function squared, is independent of the scaling angle since the
complex-scaled momentum $q e^{- i \theta}$ can be interpreted as the
change of the contour for the momentum integral of the range $[ 0, \,
\infty )$ to the rotated one with the angle $- \theta$.

An important application of the scheme developed here is to
investigate the internal structure of candidates of hadronic
molecules.  Actually, we can discuss the hadronic molecular component
of hadronic resonances in terms of the compositeness, by constructing
hadron--hadron scattering amplitudes in an effective model and
extracting the two-body wave function from the amplitudes.  In a
forthcoming paper~\cite{Sekihara:2016}, we will apply our present
scheme to the physical $N^{\ast}$ and $\Delta ^{\ast}$ resonances in a
precise $\pi N$ scattering amplitude, and discuss the meson--baryon
components for these resonances by the compositeness from the
scattering amplitude.  Finally we emphasize that, in general, the
present scheme can be applied to resonances in any other models, such
as $N^{\ast}$ and $\Delta ^{\ast}$ resonances in the dynamical
approaches of ANL--Osaka~\cite{Kamano:2013iva} and
J\"{u}lich~\cite{Ronchen:2012eg}, as long as they fully solve the
Lippmann--Schwinger equation.

\begin{acknowledgments}
  The author acknowledges M.~Oka, A.~Hosaka, H.~Nagahiro, S.~Yasui,
  H.~Kamano, T.~Hyodo, and K.~Miyahara for fruitful discussions.
  This work is partly supported by the Grants-in-Aid for Scientific
  Research from MEXT and JSPS (No.~15K17649, 
  No.~15J06538
  ).
\end{acknowledgments}

\appendix

\end{document}